\documentclass[prd,letterpaper,twocolumn,showpacs,showkeys,lengthcheck,tightenlines,
               nofootinbib,preprintnumbers,superscriptaddress]{revtex4-1}

\usepackage[utf8]{inputenc}

\usepackage{titlesec}
\usepackage{dcolumn}
\usepackage{bm}
\usepackage{amsmath,amssymb,amsfonts}

\usepackage{graphicx}
\usepackage{subfigure}
\usepackage{color}
\usepackage{bbold}

\usepackage{slashed}
\usepackage[svgnames]{xcolor}
\usepackage[english]{babel}
\usepackage{blindtext}
\usepackage{microtype}
\usepackage{tikz}

\usepackage[colorlinks=true,urlcolor=blue,citecolor=blue]{hyperref}
\usepackage{ifpdf}
\usepackage{float}

\graphicspath{{.}{figures/}}

\titlespacing{\section}{5pt}{12pt plus 4pt minus 2pt}{8pt plus 2pt minus 2pt}
\titlespacing{\subsection}{0pt}{12pt plus 4pt minus 2pt}{8pt plus 2pt minus 2pt}

\begin{document}

\title{$\rho$ meson generalized parton distributions in the Nambu--Jona-Lasinio model}

\author{Jin-Li Zhang}
\email[]{jlzhang@njnu.edu.cn}
\affiliation{Department of Physics, Nanjing Normal University, Nanjing 210023, China }

\author{Guang-Zhen Kang}
\email[]{gzkang@nju.edu.cn}
\affiliation{School of Science, Yangzhou Polytechnic Institute, Yangzhou 225127, China }

\author{Jia-Lun Ping}
\email[]{jlping@njnu.edu.cn}
\affiliation{Department of Physics, Nanjing Normal University, Nanjing 210023, China }

\begin{abstract}
In this paper, both the unpolarized and the polarized $\rho$ meson generalized parton distributions are investigated in the framework of the
Nambu--Jona-Lasinio model using proper time regularization scheme. The symmetry properties of $\rho$ meson generalized parton distributions
are checked. The three independent distribution functions in deep inelastic scattering, $F_1(x)$, $b_1(x)$ and $g_1(x)$, and
the Sachs-like charge, magnetic, and quadruple form factors $G_C(t)$, $G_M(t)$ and $G_Q(t)$, which are the first Mellin moments of
unpolarized generalized parton distributions are obtained. In addition, the $u$ quark axial vector form factors $\tilde{F}_1^u(t)$ and
$\tilde{F}_2^u(t)$ related to the axial currents from the polarized generalized parton distributions are studied. The impact parameter dependent parton distribution functions, which are the two-dimensional Fourier transform of generalized parton distributions are studied, too.
The obtained $\rho$ meson generalized parton distributions satisfy the required properties well.
\end{abstract}

\maketitle
\section{Introduction }
The understanding of hadron structure has improved greatly in recent years. However, there is still a long way to go.
An important step is the derivation of the factorization theorems, which indicates the scattering amplitudes of hard exclusive processes
are the convolution of perturbatively computable coefficient functions and generalized parton distributions (GPDs)~\cite{Mueller:1998fv,Ji:1996ek,Radyushkin:1997ki,Vanderhaeghen:1999xj,Belitsky:2001ns,Goeke:2001tz,Burkardt:2002hr,Diehl:2003ny,Belitsky:2005qn,Zhang:2020ecj,Zhang:2021mtn,Zhang:2021shm,Zhang:2021tnr}, such as deeply virtual Compton scattering (DVCS)~\cite{Radyushkin:1996nd,PhysRevD.55.7114,Collins:1998be,Belitsky:2001ns,Garcon:2002jb}, deeply virtual meson production (DVMP)~\cite{Mueller:2013caa,Favart:2015umi}, and time like Compton scattering (TCS)~\cite{Berger:2001xd,Boer:2015fwa} processes.
Quantum Chromodynamics (QCD) factorization theorems have been demonstrated for DVCS and DVMP in Refs.~\cite{Ji:1997nk,Ji:1998xh,Collins:2011zzd}.
GPDs are useful tools to shed light on the origin of nucleon spin and the three-dimensional structure of hadrons at partonic level, therefore an entirely new picture of hadrons at quark and gluon degrees of freedom is emerging. The understanding of the internal structures and
the dynamics of hadrons: the mechanical properties of angular momentum~\cite{Ji:1996ek,Ji:1996nm}, pressure, and
shear forces~\cite{Polyakov:2002yz,Polyakov:2018zvc}, is promoted with the help of GPDs.
GPDs will reduce to parton distributions functions (PDFs)~\cite{Holt:2010vj,Nguyen:2011jy,Dulat:2015mca,Ding:2019qlr,Cui:2020dlm},
form factors (FFs)~\cite{Rodriguez-Quintero:2019yec,Xu:2019ilh,Cui:2020rmu,Chen:2021guo} and parton distribution amplitudes
(PDAs)~\cite{Braun:2003wx,Radyushkin:2009zg,Shi:2014uwa,Shi:2015esa,Lu:2021sgg} in certain limits. Similar to the PDFs obtained from
inclusive deep inelastic scattering (DIS), GPDs are defined as the non-diagonal matrix elements of the same operators at a light-like
separation between the respective parton fields, which means they parameterize the quark, antiquark and gluon correlation functions involving
matrix elements between different hadron states. Due to the different momenta of the hadron states, GPDs depend not only on the
longitudinal momentum fraction $x$, but also on two other kinematic variables: the square of total momentum transfer $t$ and the skewness parameter $\xi$.

The impact parameter dependent PDFs~\cite{Burkardt:2000za,Burkardt:2000wq} are defined as the two-dimensional Fourier transform of the momentum transfer $\bm{\Delta}_{\perp}$ of GPDs in the transverse direction at zero skewness ($\xi=0$). Impact parameter dependent PDFs can be interpreted as the density of partons with longitudinal momentum fraction $x$ and transverse impact parameter $\bm{b}_{\perp}$. In the theory of GPDs, the moments of $x$ play an important role.
For $\rho$ meson, the first Mellin moments of the unpolarized GPDs are FFs associated with vector currents, the combination of these FFs related to the
Sachs-like charge, magnetic, and quadruple FFs, and the second Mellin moments correspond to the gravitational form factors. In the forward limit $t=0$, GPDs reduce to the three DIS distribution functions $F_1(x)$, $b_1(x)$ and $g_1(x)$. Furthermore, the moments of GPDs can provide
other information of nucleon structure, such as the quark orbital angular momentum~\cite{Hoodbhoy:1998yb,Ji:2016jgn} and the neutron asymmetry~\cite{Zhang:2015kna}.

GPDs of spin $1$ hadrons have been studied in Refs.~\cite{Sun:2017gtz,Cosyn:2019uaf,Cosyn:2018rdm,Cosyn:2019eeg,Berger:2001zb,Cano:2003ju,Dong:2013rk,Cosyn:2018thq,Cano:2003ju}, and spin 0
pion and kaon GPDs are calculated in Refs.~\cite{Zhang:2020ecj,Zhang:2021mtn,Zhang:2021shm}. In this work we study the $\rho$ meson GPDs~\cite{Sun:2017gtz,Sun:2018ldr,Adhikari:2018umb} in the Nambu--Jona-Lasinio (NJL)  model~\cite{RevModPhys.64.649,Buballa:2003qv,Du:2013oza,Shi:2015ufa,Lu:2016uwy,Cui:2018bor,Zhang:2018ouu,Zhang:2016zto,Cui:2017ilj,Cui:2016zqp}. NJL model has an effective Lagrangian of relativistic
fermions interacting through local fermion-fermion couplings, which can keep one of the most important fundamental symmetries of QCD, chiral symmetry,
so it is widely used in the strong interacting systems. NJL model is often taking as a good start for exploring physical phenomena where the more realistic frameworks can't yet provide insights or predictions. Our results in this work can be interpreted as the boundary values for the further evaluation.

This article is organized as follows: In Sec.~\ref{nice}, we introduce the NJL model, then give the definition and calculation of $\rho$ meson GPDs in
the NJL model. In Sec.~\ref{good}, we display the properties of $\rho$ meson GPDs, such as, the symmetry, its forward limit and the Mellin moments. A brief summary and outlook are given in Sec.~\ref{excellent}.

\section{NJL model and GPDs}\label{nice}
The $\rho$ meson, as a spin--$1$ object, is usually considered as a $q\bar{q}$ bound state. Spin--$1$ particles have more GPDs than the spin--$1/2$ ones.
In the leading twist, both the quarks and gluons have $9$ helicity conserving GPDs and $9$ helicity flip (transversity) GPDs, which were introduced in Refs.~\cite{Berger:2001zb} and ~\cite{Cosyn:2018rdm} respectively. In this work, we will calculate the $9$ helicity conserving GPDs of $u$ quark in the NJL model.

\subsection{NJL model}\label{good}
NJL model is Poincar\'{e} covariant, it describes the dynamics of interacting systems of relativistic fermions.
It is a purely fermionic theory, the gluon degrees of freedom have been integrated out. The model shows both the chiral symmetry breaking and
the dynamical mass gap generation, so it often serves as a low-energy effective theory of QCD.

The 2-flavor NJL Lagrangian is
\begin{align}\label{1}
\mathcal{L}&=\bar{\psi }\left(i\gamma ^{\mu }\partial _{\mu }-\hat{m}\right)\psi\nonumber\\
&+\frac{1}{2} G_{\pi }\left[\left(\bar{\psi }\psi\right)^2-\left( \bar{\psi }\gamma _5 \vec{\tau }\psi \right)^2\right]-\frac{1}{2}G_{\omega}\left(\bar{\psi }\gamma _{\mu}\psi\right)^2\nonumber\\
&-\frac{1}{2}G_{\rho}\left[\left(\bar{\psi }\gamma _{\mu} \vec{\tau } \psi\right)^2+\left( \bar{\psi }\gamma _{\mu}\gamma _5 \vec{\tau } \psi \right)^2\right],
\end{align}
where $\hat{m}=\text{diag}\left(m_u,m_d\right)$ is current quark mass matrix and $\vec{\tau}$ are the Pauli matrices to represent isospin.
In the limit of exact isospin symmetry, $m_u = m_d =m$. The 4-fermion coupling constants in chiral channels are labeled by $G_{\pi}$ , $G_{\omega}$,
and $G_{\rho}$.

By solving the gap equation, we obtain the dressed quark propagator in the NJL model,
\begin{align}\label{2}
S(k)=\frac{1}{{\not\!k}-M+i \varepsilon}.
\end{align}
The interaction kernel of the gap equation is local, so we obtain a constant dressed quark mass $M$, which satisfies
\begin{align}\label{3}
M=m+i  12 G_{\pi}\int \frac{\mathrm{d}^4l}{(2 \pi )^4}\text{tr}_\text{D}[S(l)],
\end{align}
where the trace is over Dirac indices. When the coupling constant is strong enough $G_{\pi} > G_{critical}$, dynamical chiral symmetry breaking can occur,
which gives a nontrivial solution $M > 0$.

NJL model is not renormalizable, so it needs a regularization method to fully define the model. We select the proper time regularization scheme~\cite{Ebert:1996vx,Hellstern:1997nv,Bentz:2001vc}.
\begin{align}\label{4}
\frac{1}{X^n}&=\frac{1}{(n-1)!}\int_0^{\infty}\mathrm{d}\tau\, \tau^{n-1}e^{-\tau X}\nonumber\\
& \rightarrow \frac{1}{(n-1)!} \int_{1/\Lambda_{\text{UV}}^2}^{1/\Lambda_{\text{IR}}^2}\mathrm{d}\tau\, \tau^{n-1}e^{-\tau X}
\end{align}
where $X$ stands for a product of propagators that have been combined using Feynman parametrization. We introduce the infrared cutoff $\Lambda_{\text{IR}}$
to mimic confinement because it is absent in the NJL model. The infrared cutoff should be of the order $\Lambda_{\text{QCD}}$ and we choose $\Lambda_{\text{IR}}=0.240$ GeV. The coupling strength $G_{\pi}$, the momentum cutoff $\Lambda_{\text{UV}}$ and the current quark mass $m$ are determined through the Gell-Mann–Oakes–Renner (GMOR) relation, $f_{\pi}^2m_{\pi}^2=-m\langle\bar{\psi}\psi \rangle$ and gap equation
$M=m-2G_{\pi}\langle\bar{\psi}\psi \rangle$, where $\langle\bar{\psi}\psi \rangle$ is two-quark condensate derived from QCD sum rules.
$m_{\pi}=0.140$ GeV is the physical pion mass, $f_{\pi}=0.092$ GeV is the pion decay constant, $m$ and $M$ are the current and the constituent quark mass,
respectively.

The self-energy of the pseudoscalar bubble diagram is
\begin{align}\label{ab35}
\Pi_{PP}(\Delta^2)
=-\frac{3 }{2\pi ^2} \int_0^1\mathrm{d}x\,\mathcal{C}_0(M^2)-\frac{3 }{4\pi ^2} \int_0^1\mathrm{d}x\, \Delta^2 \bar{\mathcal{C}}_1(\sigma_1),
\end{align}
where $\mathcal{C}_0$, $\bar{\mathcal{C}}_1$ and $\sigma_1$ are defined in Eqs. (\ref{cfun}) and (\ref{cfun1}) of the appendix.
$m_{\pi}=0.14$ GeV is coinciding with the value obtained from the pole condition $1+2G_{\pi} \Pi_{PP} (m_{\pi}^2)=0$.
$G_{\omega}$ and $G_{\rho}$ are determined by $m_{\omega}=0.782$ GeV, $m_{\rho}=0.770$ GeV through $1+2G_i \Pi_{VV} (m_i^2)=0$, $i=(\omega,\rho)$,
where $\Pi_{VV} (\Delta^2)$ is the self-energy of the vector bubble diagram defined in Eq. (\ref{vvbu}).
The parameters used in this work are given in Table \ref{tb1}.

From the inhomogeneous Bethe-Salpeter equation, the dressed quark form factors, associated with the electromagnetic currents, are~\cite{Zhang:2021shm}
\begin{align}\label{bsam}
F_{1i}(\Delta^2)=\frac{1}{1+2G_i \Pi_{VV}(\Delta^2)},\quad \quad F_{2i}(\Delta^2)=0,
\end{align}
where $i=(\omega,\rho)$, $\Pi_{VV}$ is the self-energy of the vector bubble diagram
\begin{align}\label{vvbu}
\Pi_{VV}(\Delta^2)=-\frac{3 }{\pi ^2} \int_0^1\mathrm{d}x\, x (1-x) \Delta^2 \bar{\mathcal{C}}_1(\sigma_1),
\end{align}
$\mathcal{C}_0, \bar{\mathcal{C}}_1$ are represented in terms of Gamma functions. The notations in Eqs. (\ref{cfun}) and (\ref{cfun1}) are used in the following sections.

\begin{center}
\begin{table}
\caption{Parameters used in our work. The dressed quark mass and regularization parameters are in units of GeV, while coupling constant is in units of GeV$^{-2}$, and quark condensate is in units of GeV$^3$.}\label{tb1}
\begin{tabular}{p{0.7cm} p{0.8cm} p{0.5cm} p{0.7cm}p{0.7cm}p{0.7cm}p{0.7cm}p{0.7cm}p{0.8cm}p{1.0cm}}
\hline\hline
$\Lambda_{\text{IR}}$ & $\Lambda_{\text{UV}}$ & $M$ & $m$ & $m_{\rho}$ & $G_{\pi}$ &$G_{\omega}$&$G_{\rho}$ & $Z_{\rho}$ & $\langle\bar{u}u \rangle^{1/3}$ \\
\multicolumn{5}{c}{GeV} & \multicolumn{3}{c}{GeV$^{-2}$} & GeV$^{2}$  & GeV\\
\hline
0.24 & 0.645 & 0.4 & 0.016 & 0.77 & 19.0 & 10.4 & 11.0 & 6.96 &$-$0.173\\
\hline\hline
\end{tabular}
\end{table}
\end{center}

\subsection{The definition and calculation of GPDs}\label{qq}
\begin{figure}
\centering
\includegraphics[width=0.47\textwidth]{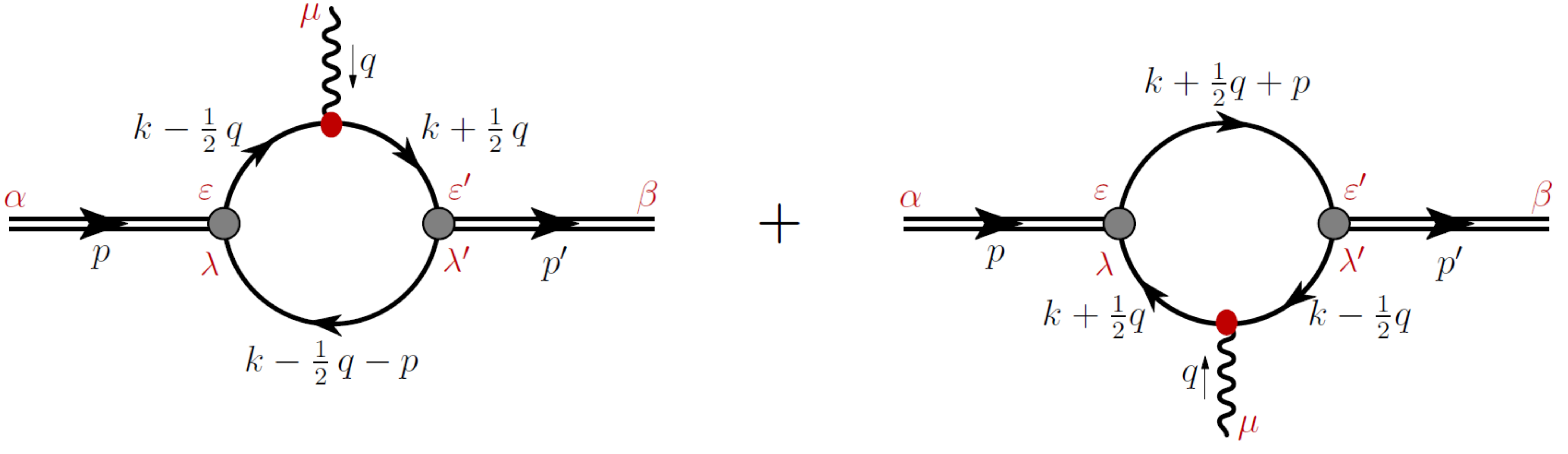}
\caption{Feynman diagrams representing the $\rho^+$ meson GPDs.}\label{GPD}
\end{figure}

The $\rho$ meson GPDs in the NJL model are illustrated in Fig. \ref{GPD}, where $p$ is the incoming and $p^{\prime}$ the outgoing $\rho$ meson momentum,
in this paper we will use the symmetry notation as Refs.~\cite{Ji:1998pc,Diehl:2003ny}, the kinematics of this process and the related quantities
are defined as
\begin{align}\label{4}
p^2=p^{\prime 2}=m_{\rho}^2, \quad \quad t=\Delta^2=(p^{\prime}-p)^2=-Q^2,
\end{align}
\begin{align}\label{5}
\xi=\frac{p^+ - p^{\prime +}}{p^+ +p^{\prime +}},\quad P=\frac{p+p^{\prime}}{2},\quad n^2=0,
\end{align}
$\xi$ is the skewness parameter, in the light-cone coordinate
\begin{align}\label{4A}
v^{\pm}=(v^0\pm v^3), \quad  \mathbf{v}=(v^1,v^2),
\end{align}
for any four-vector, $n$ is the light-cone four-vector defined as $n=(1,0,0,-1)$, in the light-cone coordinate
\begin{align}\label{4B}
v^+=v\cdot n.
\end{align}

The unpolarized and polarized GPDs are defined through the two-parton correlation functions for quarks as
\begin{subequations}\label{rhoc}
\begin{align}
V_{\lambda^{\prime}\lambda}^q&=\frac{1}{2}\int \frac{\mathrm{d}z}{2\pi}e^{ix z P^+} \langle p^{\prime},\lambda^{\prime}
|\bar{q}(-\frac{1}{2}z)\gamma^+q(\frac{1}{2}z)|p,\lambda\rangle\nonumber\\
&=\epsilon^{\prime *\mu}V_{\mu\nu}^q \epsilon_{\nu}\,, \\
A_{\lambda^{\prime}\lambda}^q&=\frac{1}{2}\int \frac{\mathrm{d}z}{2\pi}e^{ix z P^+} \langle p^{\prime},\lambda^{\prime}|\bar{q}(-\frac{1}{2}z)\gamma^+\gamma^5 q(\frac{1}{2}z)|p,\lambda\rangle\nonumber\\
&=\epsilon^{\prime *\mu}A_{\mu\nu}^q \epsilon_{\nu},
\end{align}
\end{subequations}
where $\epsilon=\epsilon(p,\lambda)$ [or $\epsilon^{\prime}=\epsilon^{\prime}(p^{\prime},\lambda^{\prime})$] and
$\lambda$ (or $\lambda^{\prime}$) $= 0, \pm 1$ are the initial (or final) polarization vector and its helicity. How many numbers of independent
Lorentz structures the correlators can be decomposed depends on the hadron spin. The helicity counting rule restricts that there are in total $9$ helicity conserving GPDs for spin-$1$ particles, including $5$ unpolarized and $4$ polarized. Tensors $V_{\mu\nu}^q$ depend on four vectors $p$, $p^{\prime}$
and $n$. Due to the orthogonality conditions of the polarization vector $\epsilon\cdot p =\epsilon' \cdot p^{\prime} =0$, one can keep tensors
which do not vanish when contracted with $\epsilon_{\nu}$, $\epsilon_{\mu}^{\prime}$. These were first proposed in Ref.~\cite{Berger:2001zb}, but there are some differences in the formulas because our ``P'' here is half of the ``P'' in their definition.

The five tensor structures of $V_{\mu\nu}^q$ are
\begin{align}\label{a10}
\{g^{\mu\nu},\, P^{\nu}n^{\mu},\, P^{\mu}n^{\nu},\,P^{\mu}P^{\mu},\, n^{\mu}n^{\nu}\},
\end{align}
similarly, $A_{\mu\nu}^q$ are composed of seven tensors
\begin{align}
\{&\epsilon_{\alpha \beta \mu\nu } p^{\alpha}p^{\prime \beta }, \quad \epsilon_{\alpha \beta \mu\nu } n^{\alpha}p^{\beta }, \quad \epsilon_{\alpha \beta \mu\nu } n^{\alpha}p^{\prime \beta },\nonumber\\
& \epsilon_{\alpha \beta \rho \mu }p^{\alpha}p^{\prime \beta }n^{\rho}n^{\nu}, \quad \epsilon_{\alpha \beta \rho \mu }p^{\alpha}p^{\prime \beta }n^{\rho}p^{\prime \nu},\nonumber\\
& \epsilon_{\alpha \beta \rho \nu }p^{\alpha}p^{\prime \beta }n^{\rho}n^{\mu},  \quad \epsilon_{\alpha \beta \rho \nu }p^{\alpha}p^{\prime \beta }n^{\rho}p^{\mu} \},
\end{align}
where $\epsilon_{0123 }=1$. One can find that only four of these seven are linearly independent due to the Schouten identities. Consequently, the vector and axial vector quark correlators can be decomposed as follows:
\begin{widetext}
\begin{align}\label{aX1}
V_{\mu\nu}^q&=-g^{\mu\nu}H_1^q+\frac{n^{\mu}P^{\nu}+P^{\mu}n^{\nu}}{n\cdot P}H_2^q-\frac{2P^{\mu}P^{\nu}}{m_{\rho}^2}H_3^q+\frac{n^{\mu}P^{\nu}-P^{\mu}n^{\nu}}{n\cdot P}H_4^q+\left[m_{\rho}^2\frac{n^{\mu}n^{\nu}}{(n\cdot P)^2}+\frac{1}{3}g^{\mu\nu}\right]H_5^q,
\end{align}
\begin{align}\label{aX2}
A_{\mu\nu}^q&=- i\frac{\epsilon_{+\mu\nu\alpha} P^{\alpha}}{P\cdot n}\tilde{H}_1^q+ \frac{2i \Delta^{\alpha}P^{\beta}}{P\cdot n}\frac{\epsilon_{+\alpha\beta \nu}P^{\mu}+\epsilon_{+\alpha\beta \mu}P^{\nu}}{m_{\rho}^2}\tilde{H}_2^q+ \frac{2i \Delta^{\alpha}P^{\beta}}{P\cdot n}\frac{\epsilon_{+\alpha\beta \nu}P^{\mu}-\epsilon_{+\alpha\beta \mu}P^{\nu}}{m_{\rho}^2}\tilde{H}_3^q\nonumber\\
&+\frac{i \Delta^{\alpha}P^{\beta}}{P\cdot n}\frac{\epsilon_{+\alpha\beta \nu}n^{\mu}+\epsilon_{+\alpha\beta \mu}n^{\nu}}{2P\cdot n}\tilde{H}_4^q,
\end{align}
\end{widetext}
where $H_i^q$ represents $H_i^q(x,\xi,t)$. Here we only calculate the $u$ quark GPDs of the $\rho^+$ meson. In the NJL model, for $\rho^+$ meson, the quark contents are $u\bar{d}$, the GPDs of $\bar{d}$ quark can be obtained from $u$ quark GPDs through symmetry property. We will use $\rho $ to represent $\rho^+$, omitting the superscript hereafter when there is no ambiguity.

The operators in Fig. \ref{GPD}
\begin{subequations}\label{a91}
\begin{align}\label{6A}
\textcolor{red}{\bullet}_1 &=\gamma^+\delta(x-\frac{k^+}{P^+})\,, \\
\textcolor{red}{\bullet}_2 &=\gamma^+\gamma^5 \delta(x-\frac{k^+}{P^+}),
\end{align}
\end{subequations}
$\textcolor{red}{\bullet}_1$ stands for the unpolarized GPDs and $\textcolor{red}{\bullet}_2$ stands for the polarized GPDs.
The $\rho$ meson vertex function, in the light-cone normalization, is defined as
\begin{align}\label{6C}
\Gamma_{\rho}^{\mu}=\sqrt{Z_{\rho}}\gamma^{\mu},
\end{align}
where $Z_{\rho}$ is the square of the effective meson-quark-quark coupling constant, which is defined as 
\begin{align}\label{ab35}
Z_{\rho}^{-1}&=-\frac{\partial}{\partial \Delta^2}\Pi_{VV}(\Delta^2)|_{\Delta^2=m_{\rho}^2}.
\end{align}
In the NJL model, $u$ quark GPDs of $\rho$ meson are defined as
\begin{align}\label{gpddd}
V_{\mu\nu}^u&=2i N_c Z_{\rho}\int \frac{\mathrm{d}^4k}{(2 \pi )^4}\delta_n^x (k)\nonumber\\
&\times  \text{tr}_{\text{D}}\left[\gamma^{\mu} S \left(k_{+\Delta}\right)\gamma^+S\left(k_{-\Delta}\right)\gamma^{\nu} S\left(k-P\right)\right],
\end{align}
\begin{align}\label{tgpddd}
A_{\mu\nu}^u&=2i N_c Z_{\rho}\int \frac{\mathrm{d}^4k}{(2 \pi )^4}\delta_n^x (k)\nonumber\\
&\times  \text{tr}_{\text{D}}\left[\gamma^{\mu} S \left(k_{+\Delta}\right)\gamma^+\gamma^5 S\left(k_{-\Delta}\right)\gamma^{\nu} S\left(k-P\right)\right],
\end{align}
where $\text{tr}_{\text{D}}$ indicates a trace over spinor indices, $\delta_n^x (k)=\delta (xP^+-k^+)$, $k_{+\Delta}=k+\frac{\Delta}{2}$, $k_{-\Delta}=k-\frac{\Delta}{2}$.
With the help of the following formulas ($D(k^2)=k^2-M^2$)
\begin{subequations}\label{dr}
\begin{align}
p\cdot \Delta&=-\frac{\Delta^2}{2}\,, \\
k\cdot \Delta&=\frac{1}{2} \left(D(k_{+\Delta}^2)-D(k_{-\Delta}^2)\right)\,, \\
k\cdot p&=-\frac{1}{2} \left(D((k-P)^2)-D(k_{-\Delta}^2)-m_{\rho}^2+\frac{\Delta^2}{2}\right)\,, \\
k^2&=\frac{1}{2} \left(D(k_{+\Delta}^2)+D(k_{-\Delta}^2)\right)+M^2-\frac{\Delta^2}{4},
\end{align}
\end{subequations}
and Feynman parameterizations, various GPDs of $u$ quark can be obtained from Eqs. (\ref{gpddd}) and (\ref{tgpddd}),
\begin{align}\label{ugpd1}
H_1^u&=\frac{N_cZ_{\rho}}{8\pi^2} \left( \theta_{\bar{\xi} 1} \bar{\mathcal{C}}_1(\sigma_2)+ \theta_{\xi 1} \bar{\mathcal{C}}_1(\sigma_3)+\theta_{\bar{\xi} \xi}  \frac{x}{\xi} \bar{\mathcal{C}}_1(\sigma_4)\right)\nonumber\\
&-\frac{N_cZ_{\rho} }{4\pi^2}\int_0^1  \mathrm{d}\alpha \, \theta_{\alpha \xi} \frac{x}{\xi}  \bar{\mathcal{C}}_1(\sigma_5)\nonumber\\
&+\frac{N_cZ_{\rho} }{8\pi^2}\int_0^1  \mathrm{d}\alpha  \frac{\theta_{\alpha \xi}}{\xi} (2\,x\,m_{\rho}^2 -t (x-1)) \frac{\bar{\mathcal{C}}_2(\sigma_5)}{\sigma_5},
\end{align}
\begin{align}\label{ugpd2}
H_2^u&=\frac{N_cZ_{\rho}}{8\pi^2}  \frac{\theta_{\bar{\xi} \xi}}{\xi} \bar{\mathcal{C}}_1(\sigma_4)\nonumber\\
&+\frac{N_cZ_{\rho}}{8\pi^2} \left(\frac{\theta_{\bar{\xi} 1}}{(1-\xi)} \bar{\mathcal{C}}_1(\sigma_2)+\frac{\theta_{\xi 1} }{(1+\xi)} \bar{\mathcal{C}}_1(\sigma_3)\right)\nonumber\\
&+\frac{N_cZ_{\rho} }{4\pi^2}\int_0^1  \mathrm{d}\alpha \frac{\theta_{\alpha \xi}}{\xi}(\alpha -1)\bar{\mathcal{C}}_1(\sigma_5)\nonumber\\
&+\frac{N_cZ_{\rho} }{8\pi^2}\int_0^1  \mathrm{d}\alpha \frac{\theta_{\alpha \xi}}{\xi}\left(2\,m_{\rho}^2+t(1-\alpha)\right)\frac{\bar{\mathcal{C}}_2(\sigma_5)}{\sigma_5},
\end{align}
\begin{align}\label{ugpd3}
H_3^u&=-m_{\rho}^2 \frac{N_cZ_{\rho}}{2\pi^2}\int_0^1  \mathrm{d}\alpha  \frac{\theta_{\alpha \xi}}{\xi}\nonumber\\
&\times \left(x\left(\frac{x-\alpha}{\xi}\right)^2-x(\alpha-1)^2\right)\frac{\bar{\mathcal{C}}_2(\sigma_5)}{\sigma_5},
\end{align}
\begin{align}\label{pgpd1}
\tilde{H}_1^u&=\frac{N_cZ_{\rho}}{8\pi ^2} \left(\frac{\theta_{\bar{\xi} 1}}{(1+\xi) } \bar{\mathcal{C}}_1(\sigma_2)+\frac{\theta_{\xi 1}}{(1-\xi)}\bar{\mathcal{C}}_1(\sigma_3)\right)\nonumber\\
&-\frac{N_cZ_{\rho}}{4\pi^2} \int_0^1  \mathrm{d}\alpha\, \frac{\theta_{\alpha \xi}}{ \xi} \bar{\mathcal{C}}_1(\sigma_5)\nonumber\\
&+\frac{N_cZ_{\rho}}{8\pi^2} \int_0^1  \mathrm{d}\alpha\, \frac{\theta_{\alpha \xi}}{\xi} (2\alpha \,m_{\rho}^2+t(1-\alpha))\, \frac{\bar{\mathcal{C}}_2(\sigma_5)}{\sigma_5},
\end{align}

\begin{align}\label{pgpd2}
\tilde{H}_2^u&=-m_{\rho}^2\frac{N_cZ_{\rho}}{8\pi^2}\int_0^1  \mathrm{d}\alpha \, \frac{\theta_{\alpha \xi} }{\xi}\nonumber\\
&\times \left((1-\alpha)^2 -\left(\frac{x-\alpha}{\xi}\right)^2\right) \frac{\bar{\mathcal{C}}_2(\sigma_5)}{\sigma_5},
\end{align}
and
\begin{subequations}\label{region1}
\begin{align}
\theta_{\bar{\xi} 1}&=x\in[-\xi, 1]\,, \\
\theta_{\xi 1}&=x\in[\xi, 1]\,, \\
\theta_{\bar{\xi} \xi}&=x\in[-\xi, \xi]\,, \\
\theta_{\alpha \xi}&=x\in[\alpha (\xi +1)-\xi , \alpha  (1-\xi)+\xi ]\cap x\in[-1,1],
\end{align}
\end{subequations}
where $x$ only exist in the corresponding region, all the values of the $\theta$ functions are $1$ in the corresponding region and $0$ out of it. One can write $\theta_{\bar{\xi} \xi}/\xi=\Theta(1-x^2/\xi^2) $, where $\Theta(x)$ is the Heaviside function, and
$\theta_{\alpha \xi}/\xi=\Theta((1-\alpha)^2-(x-\alpha)^2/
\xi^2)\Theta(1-x^2)$. These results are in the region $\xi > 0$. Under the transformation $\xi \rightarrow -\xi $,
$\theta_{\bar{\xi} 1} \leftrightarrow \theta_{\xi 1}$ and $\theta_{\bar{\xi} \xi}/\xi$, $\theta_{\alpha \xi}/\xi$ are invariant. The GPDs obtained satisfy the required properties, which will be checked in the next section.

\begin{figure*}
\centering
\includegraphics[width=0.47\textwidth]{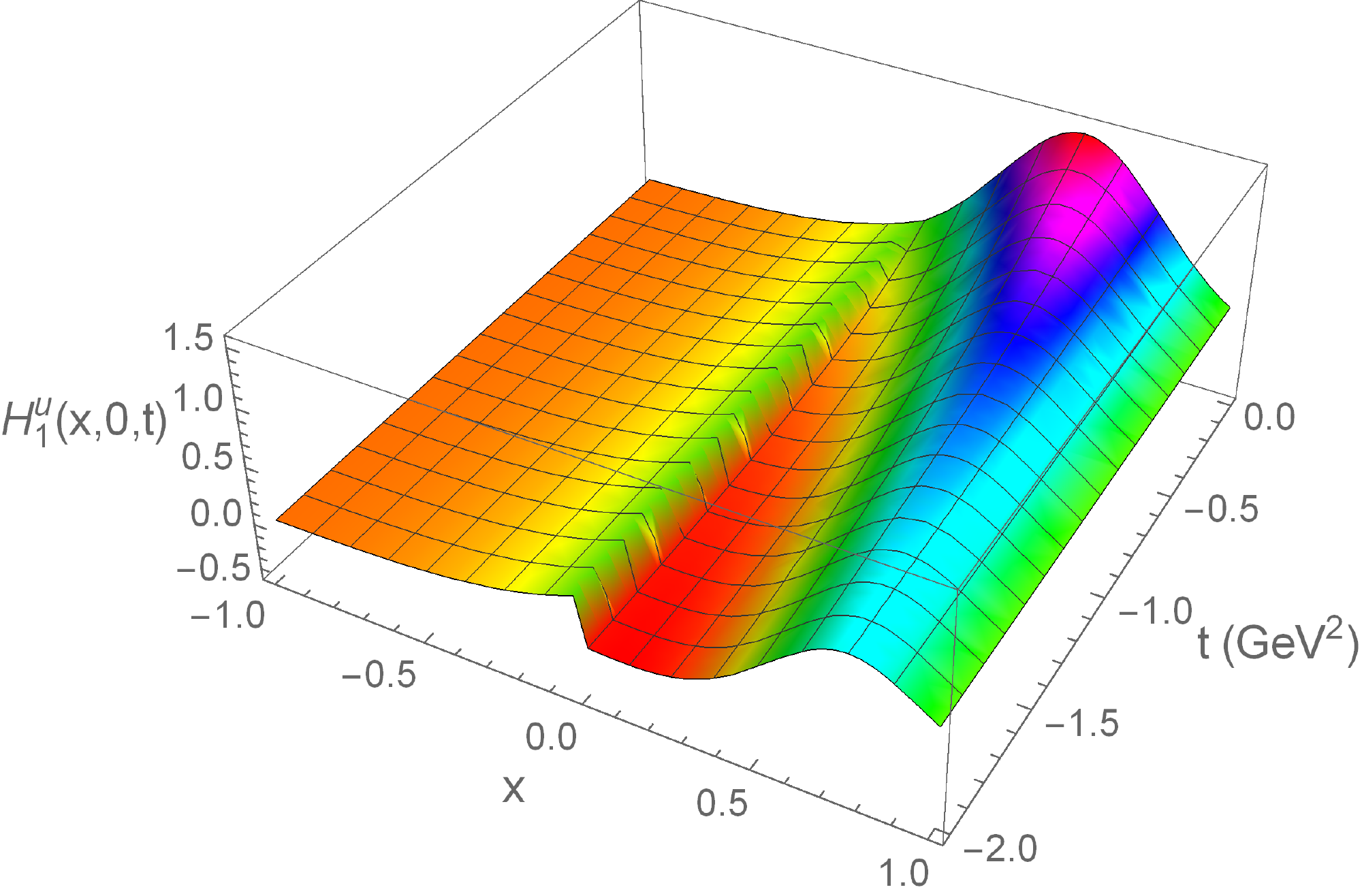}
\qquad
\includegraphics[width=0.47\textwidth]{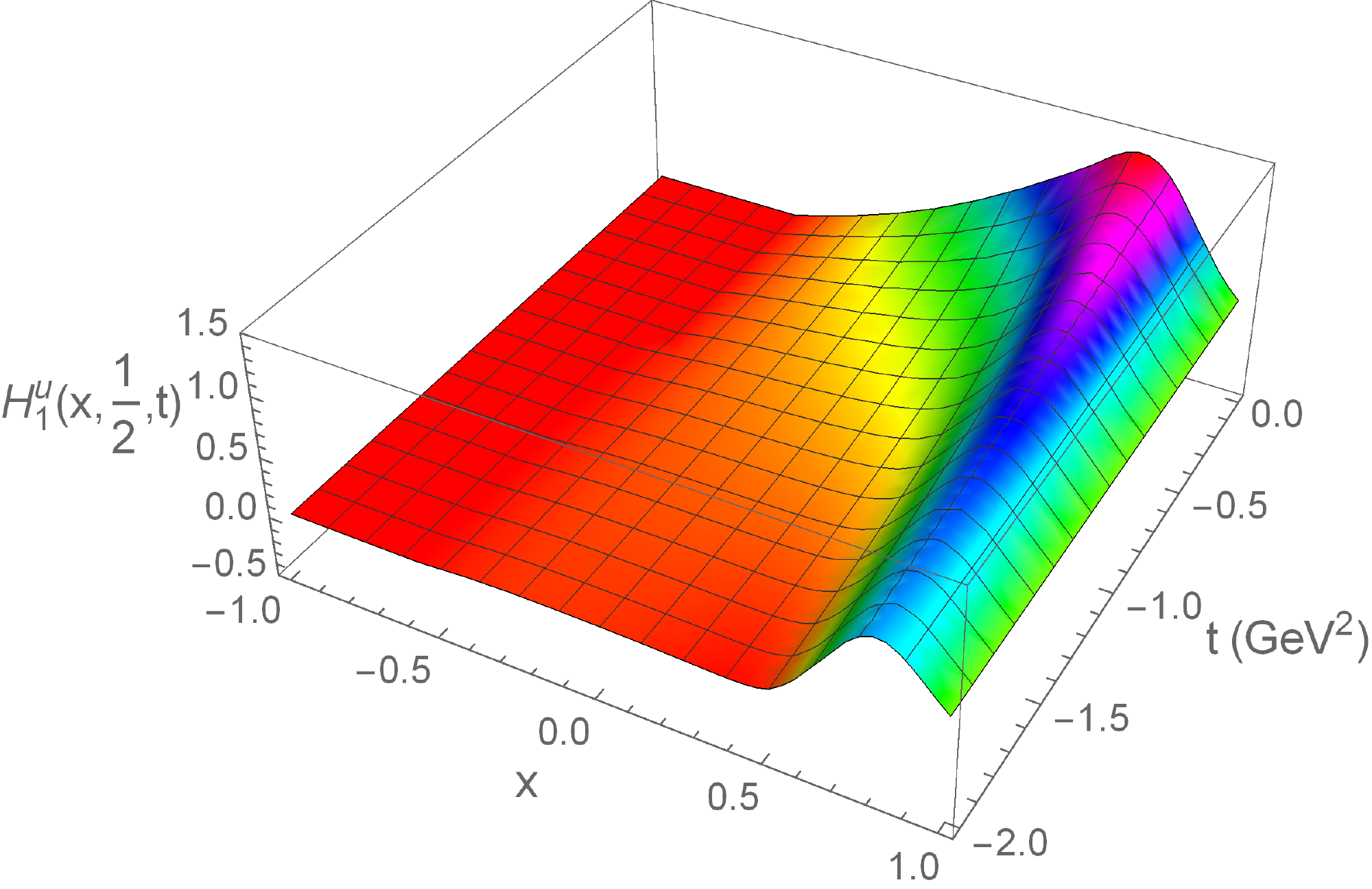}
\caption{The unpolarized $u$ quark GPDs of $\rho$ meson. \emph{left panel}: $H_1^u(x,0,t)$; \emph{right panel}: $H_1^u(x,0.5,t)$.}\label{h1}
\end{figure*}

\begin{figure*}
\centering
\includegraphics[width=0.47\textwidth]{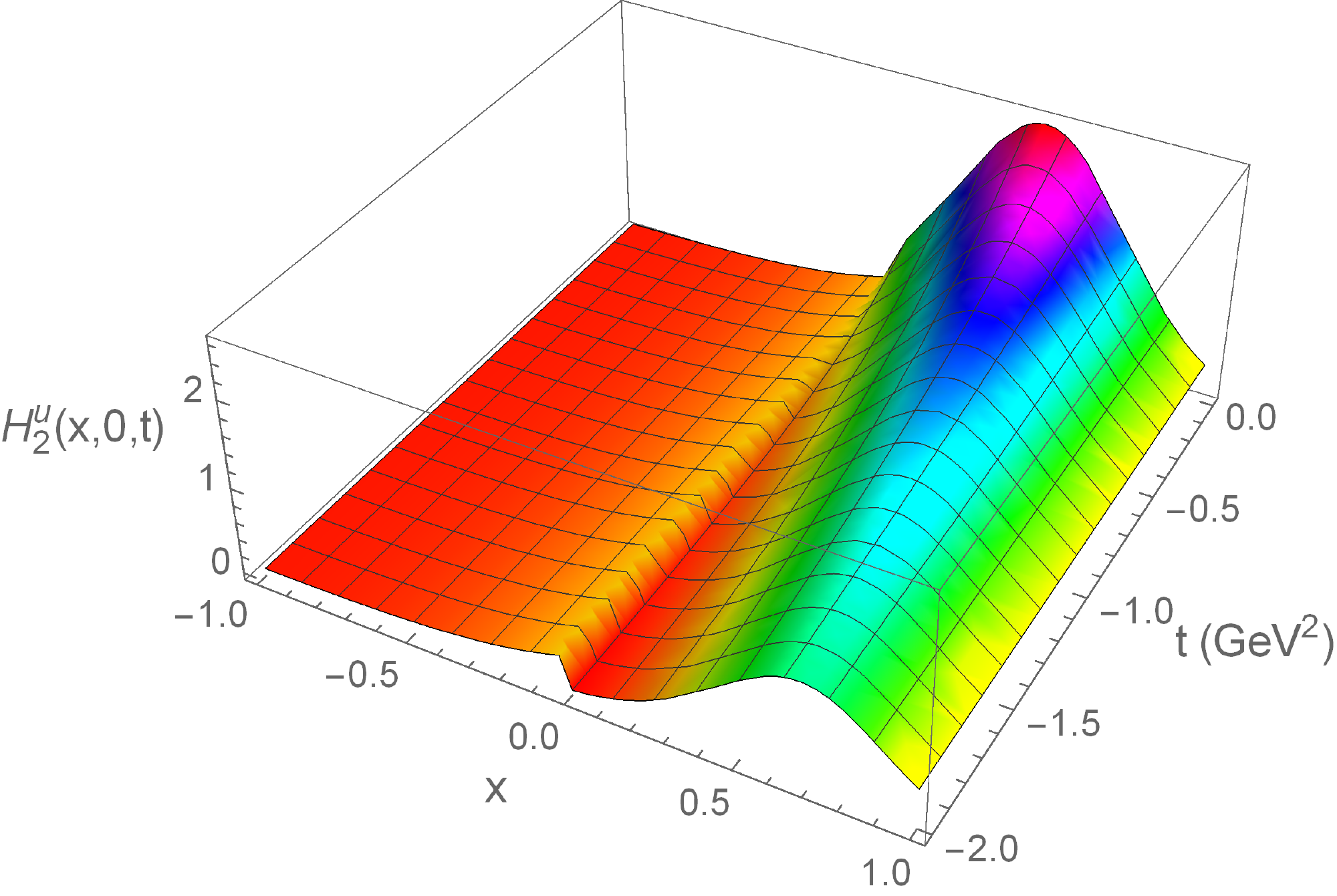}
\qquad
\includegraphics[width=0.47\textwidth]{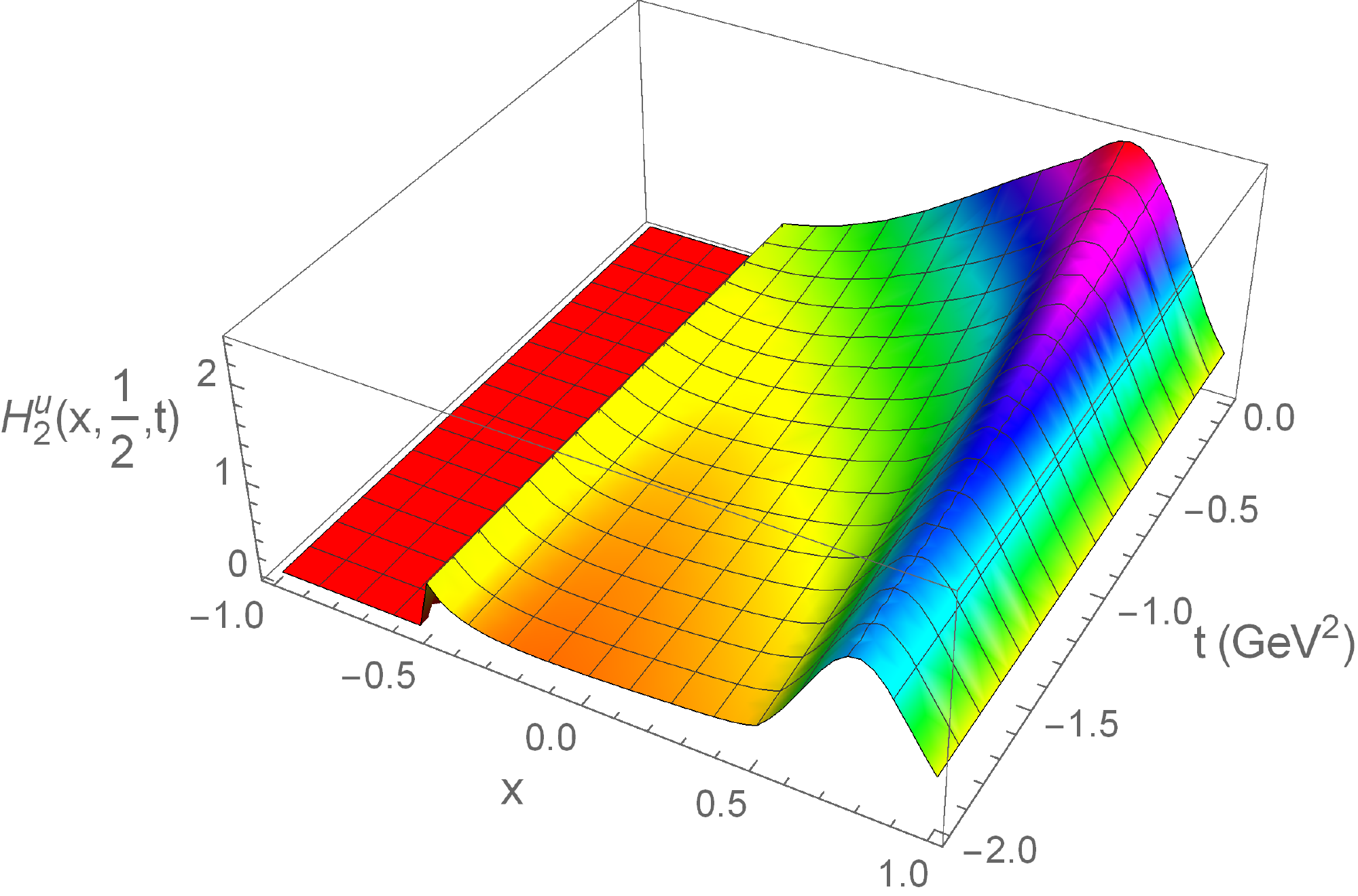}
\caption{The unpolarized $u$ quark GPDs of $\rho$ meson. \emph{left panel}: $H_2^u(x,0,t)$; \emph{right panel}: $H_2^u(x,0.5,t)$.}\label{h2}
\end{figure*}

\begin{figure*}
\centering
\includegraphics[width=0.47\textwidth]{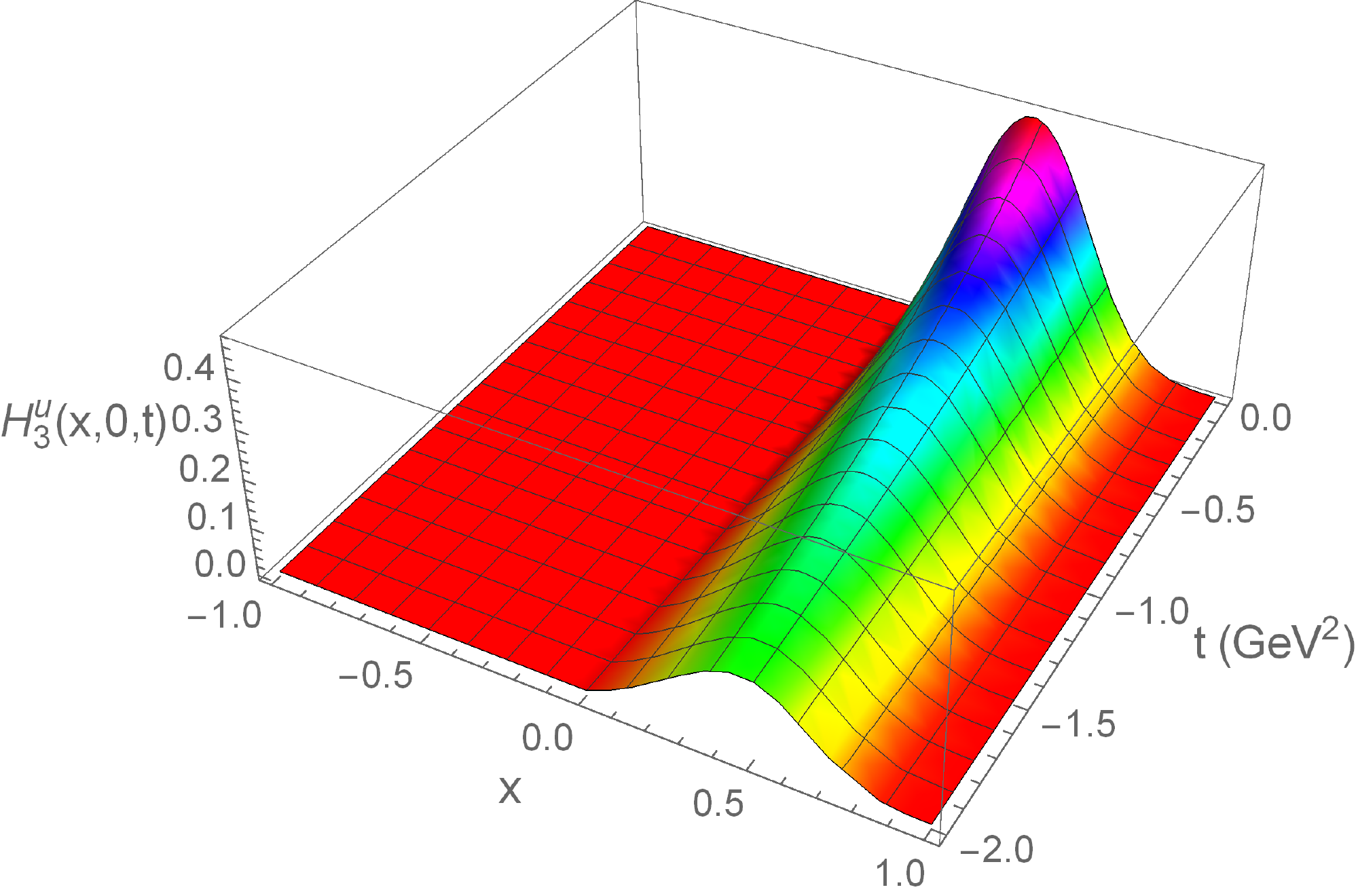}
\qquad
\includegraphics[width=0.47\textwidth]{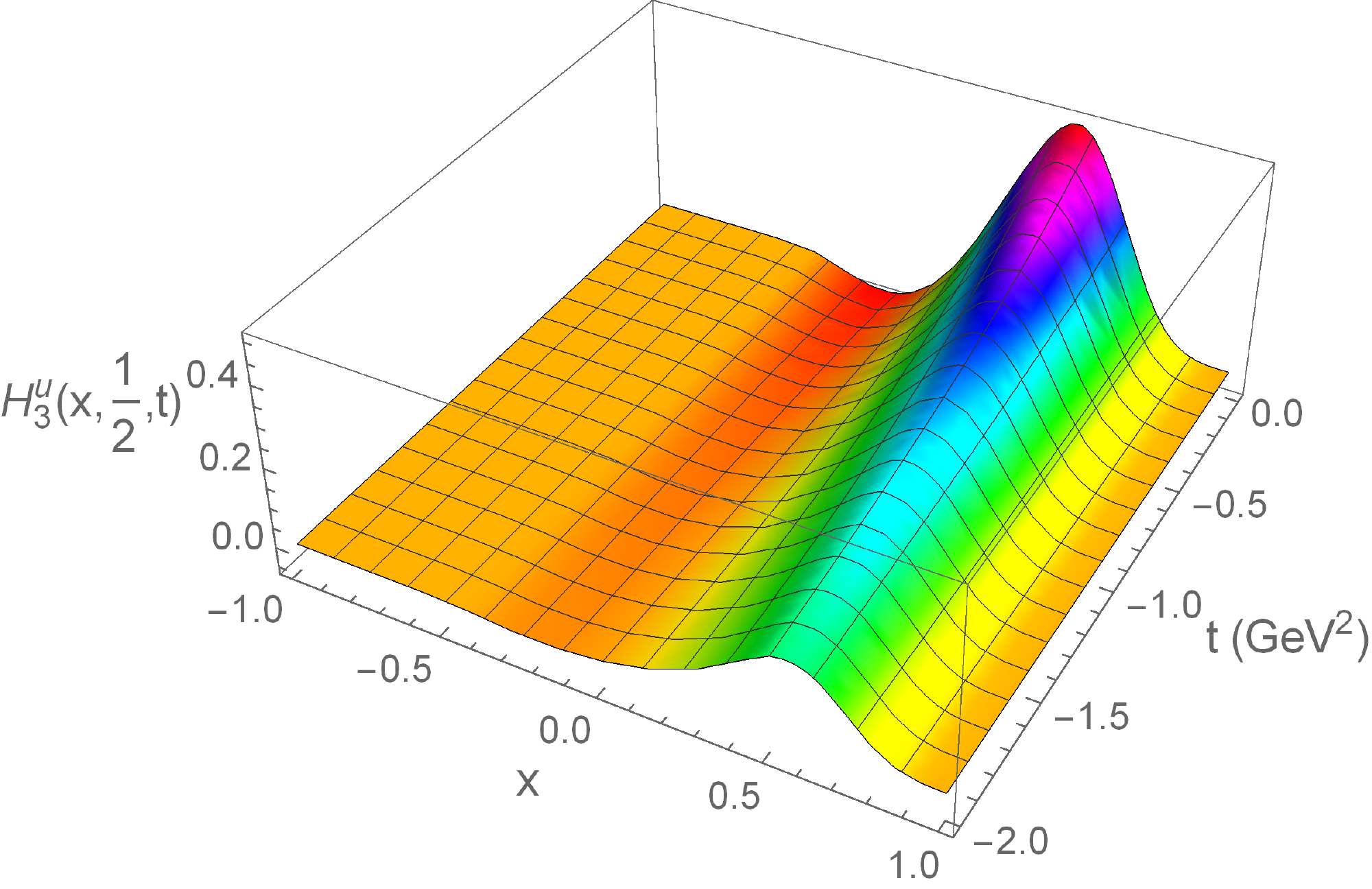}
\caption{The unpolarized $u$ quark GPDs of $\rho$ meson. \emph{left panel}: $H_3^u(x,0,t)$; \emph{right panel}: $H_3^u(x,0.5,t)$.}\label{h3}
\end{figure*}

\begin{figure*}
\centering
\includegraphics[width=0.47\textwidth]{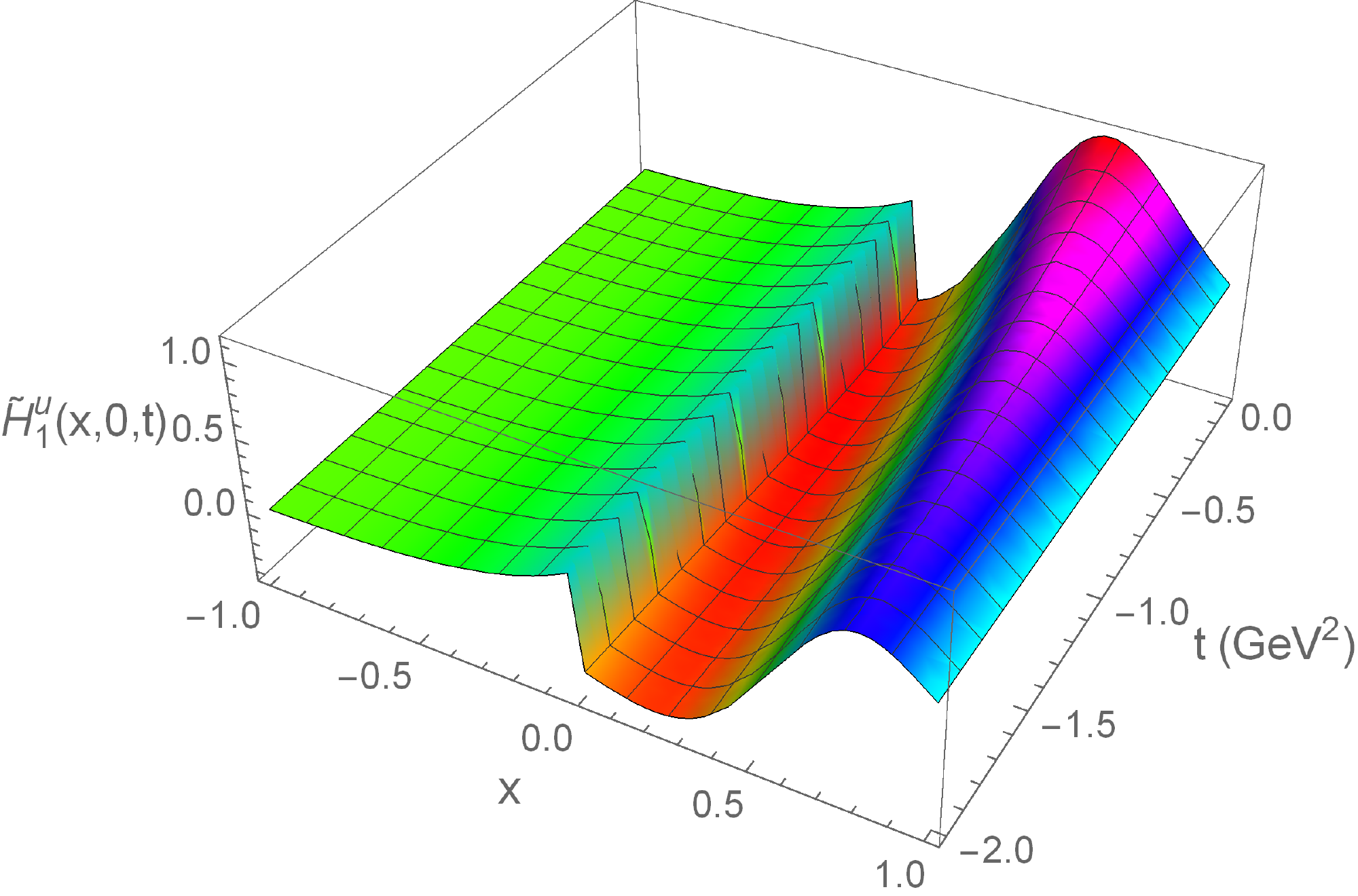}
\qquad
\includegraphics[width=0.47\textwidth]{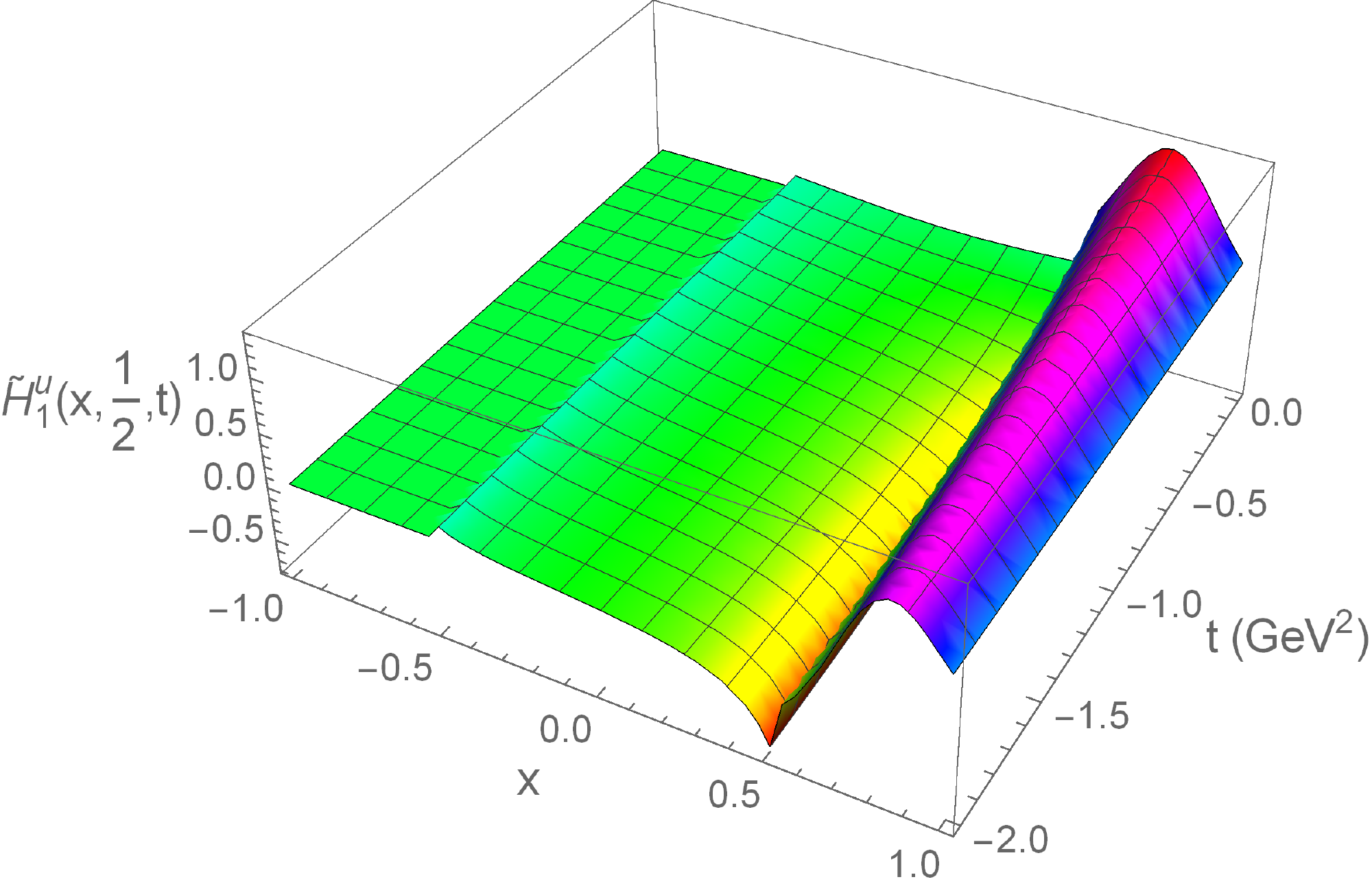}
\caption{The polarized $u$ quark GPDs of $\rho$ meson. \emph{left panel}: $\tilde{H}_1^u(x,0,t)$; \emph{right panel}: $\tilde{H}_1^u(x,0.5,t)$.}\label{th1}
\end{figure*}

\begin{figure*}
\centering
\includegraphics[width=0.47\textwidth]{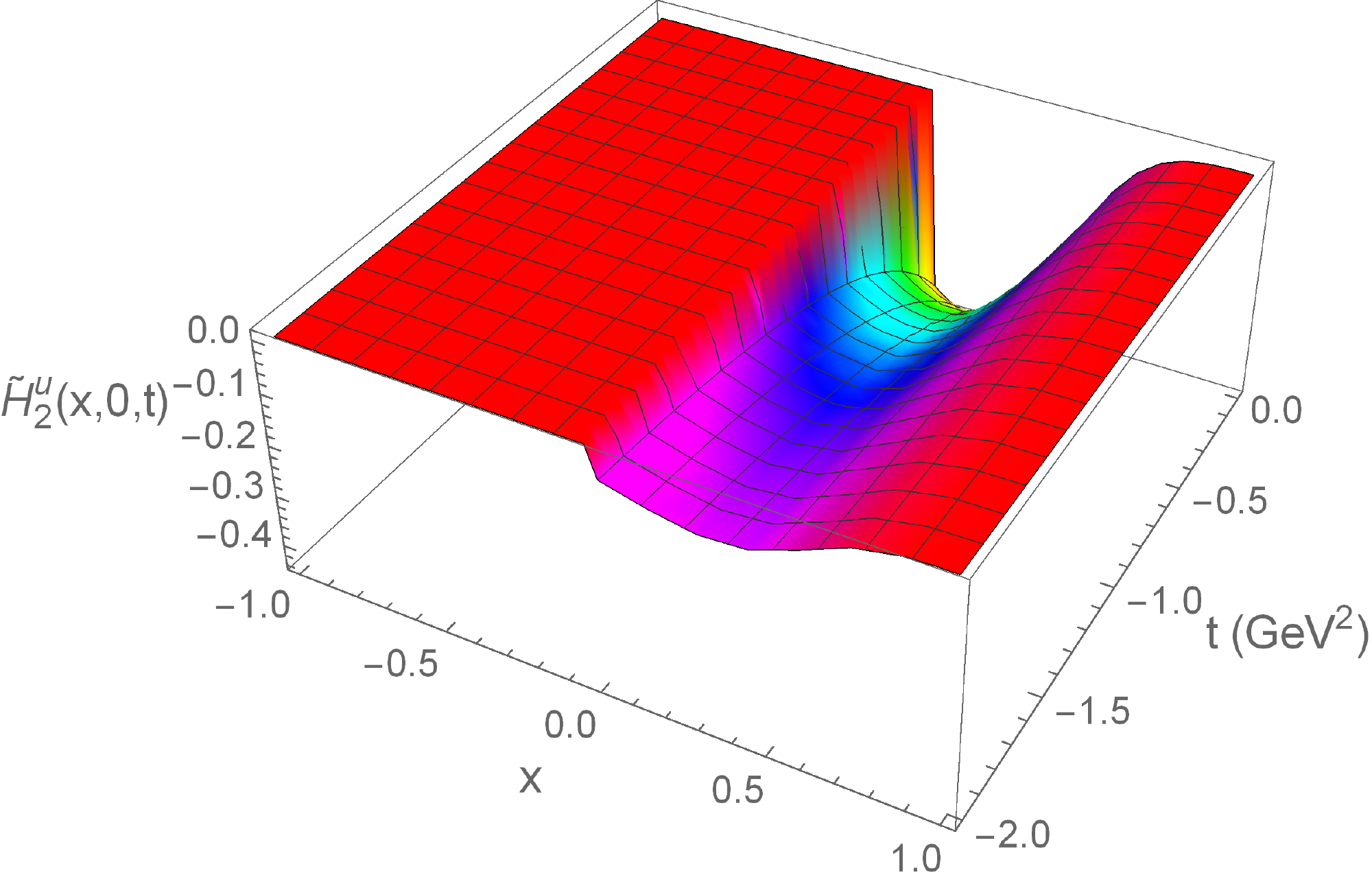}
\qquad
\includegraphics[width=0.47\textwidth]{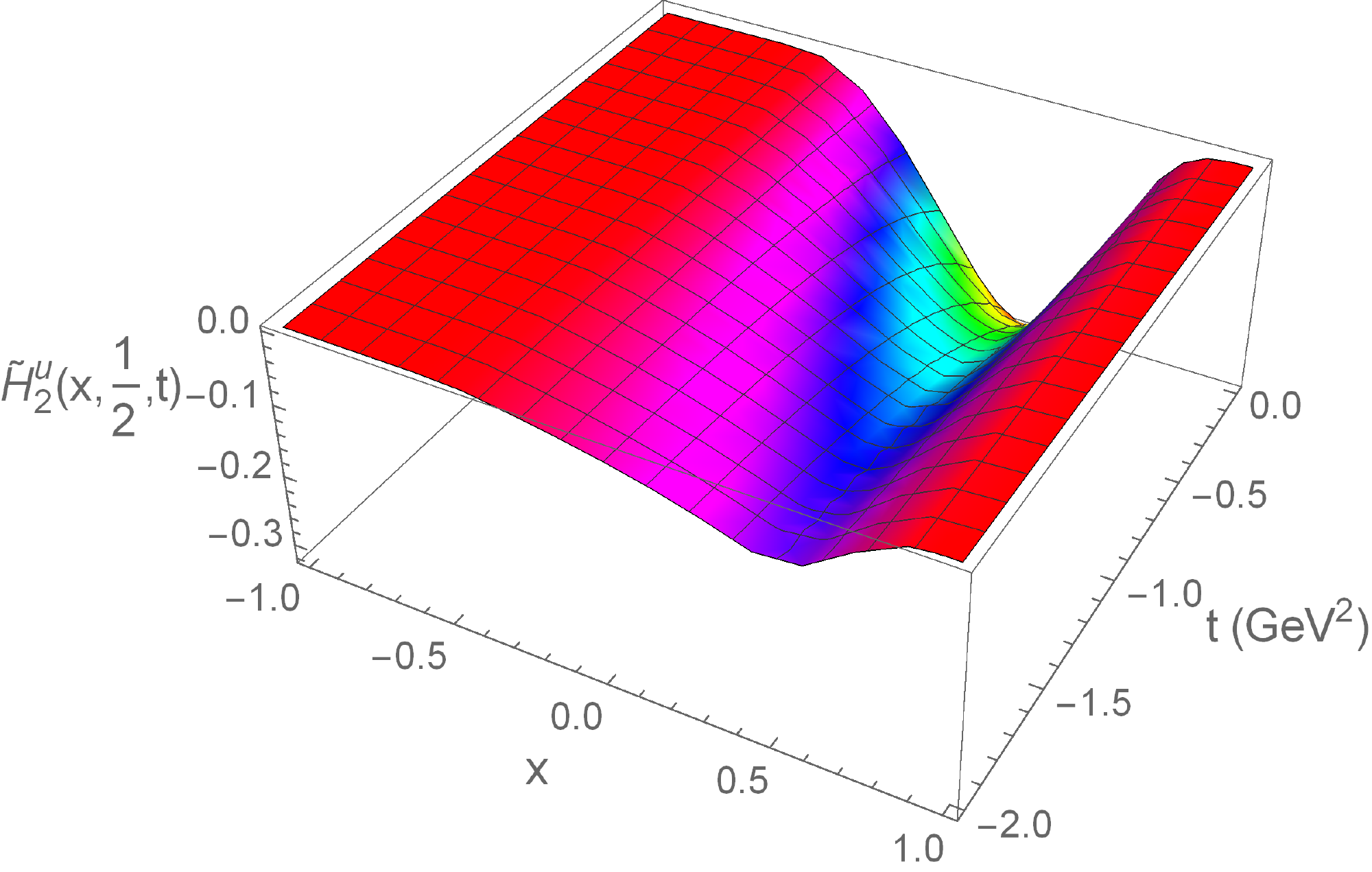}
\caption{The polarized $u$ quark GPDs of $\rho$ meson. \emph{left panel}: $\tilde{H}_2^u(x,0,t)$; \emph{right panel}: $\tilde{H}_2^u(x,0.5,t)$.}\label{th2}
\end{figure*}

\begin{figure*}
\centering
\includegraphics[width=0.47\textwidth]{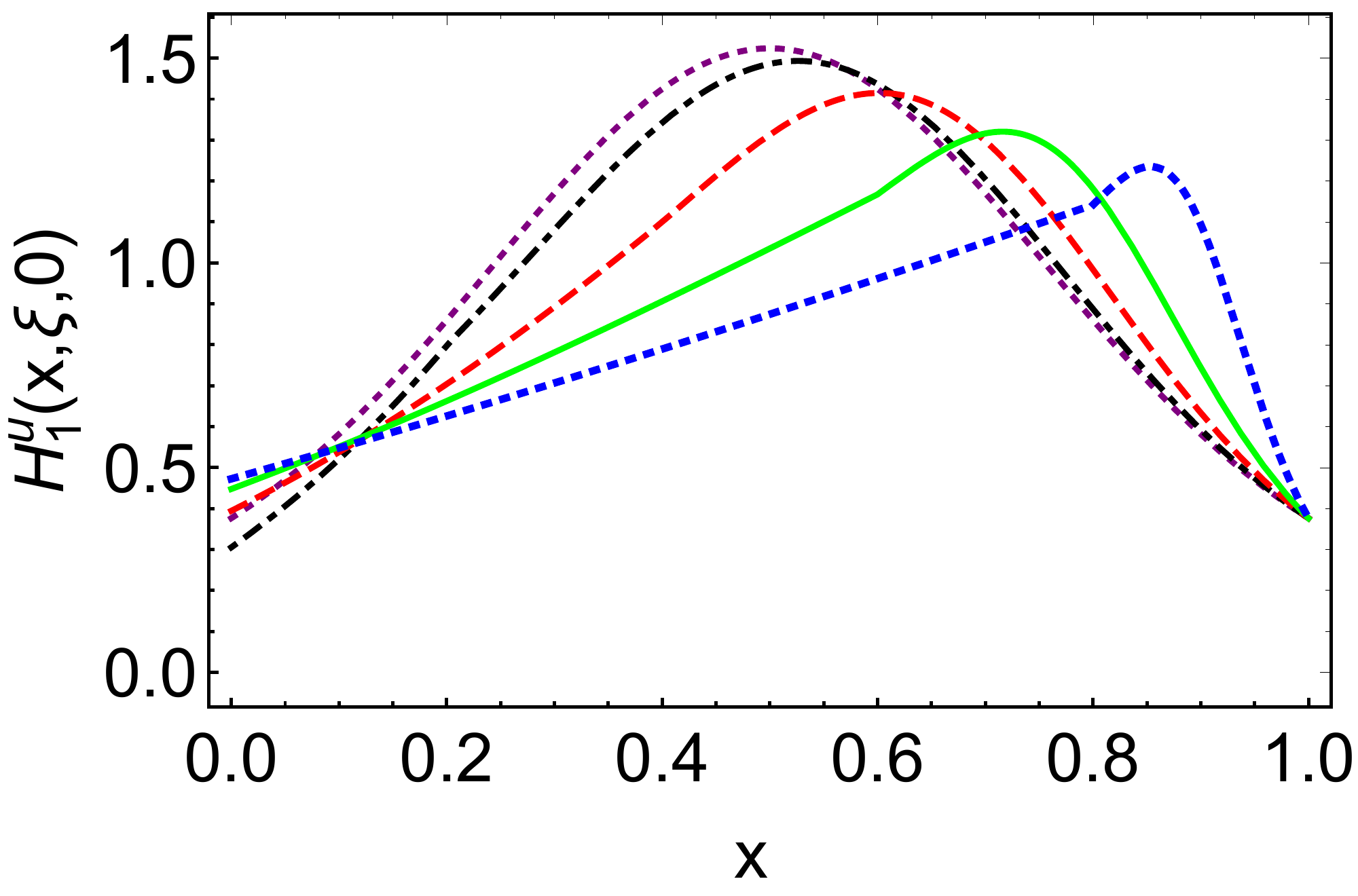}
\qquad
\includegraphics[width=0.47\textwidth]{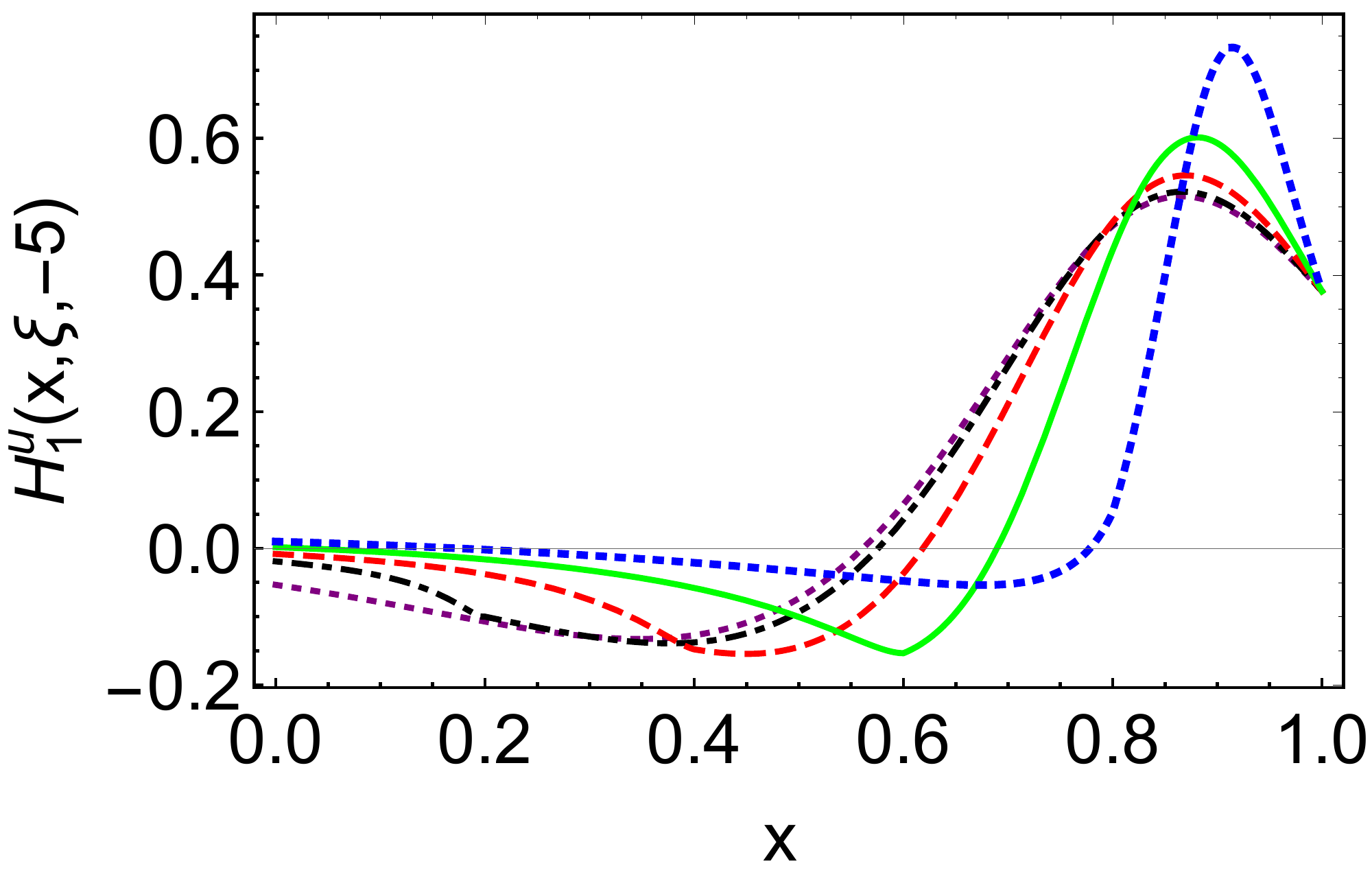}
\caption{$H_1^u(x,\xi,t)$ at $t=0$ GeV$^2$ (left panel) and $t=-5$ GeV$^2$ (right panel). $\xi=0$ (dotted purple), 0.2 (dot-dashed black),
 0.4 (dashed red), 0.6 (solid green), 0.8 (thick-dotted blue). }\label{h11} 
\end{figure*}

\begin{figure*}
\centering
\includegraphics[width=0.47\textwidth]{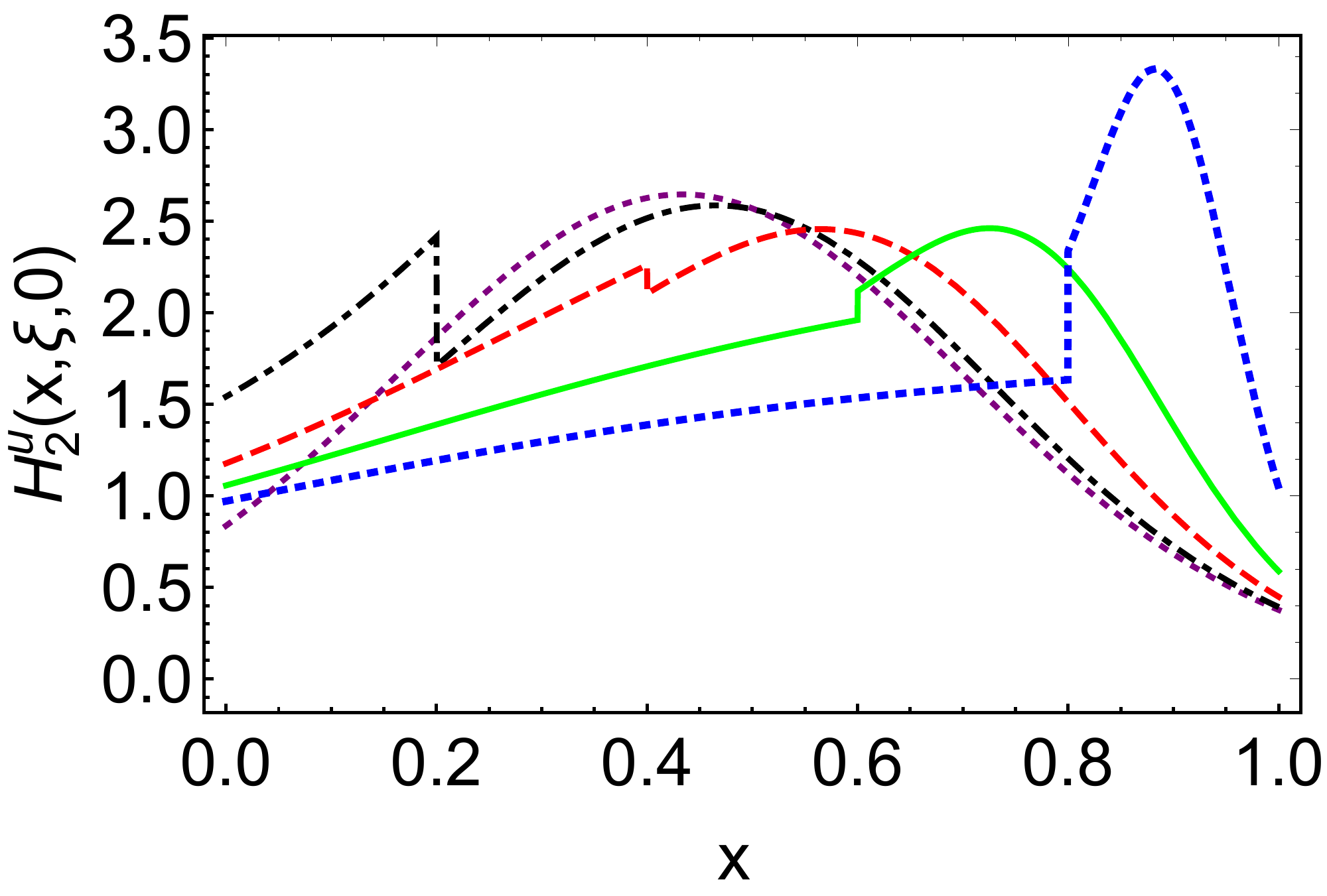}
\qquad
\includegraphics[width=0.47\textwidth]{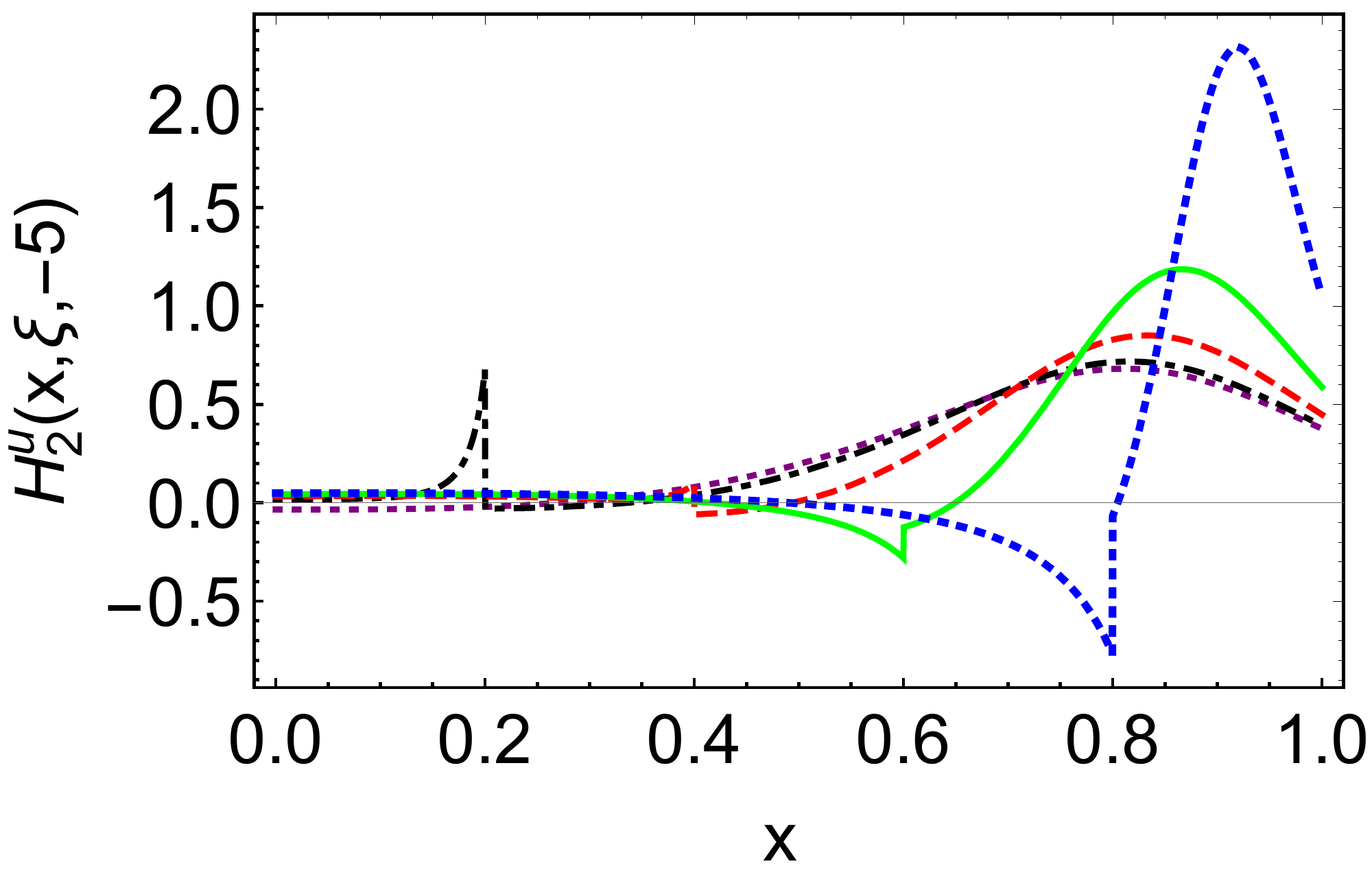}
\caption{$H_2^u(x,\xi,t)$ at $t=0$ GeV$^2$ (left panel) and $t=-5$ GeV$^2$ (right panel). $\xi=0$ (dotted purple), 0.2 (dot-dashed black),
 0.4 (dashed red), 0.6 (solid green), 0.8 (thick-dotted blue). }\label{h12} 
\end{figure*}

\begin{figure*}
\centering
\includegraphics[width=0.47\textwidth]{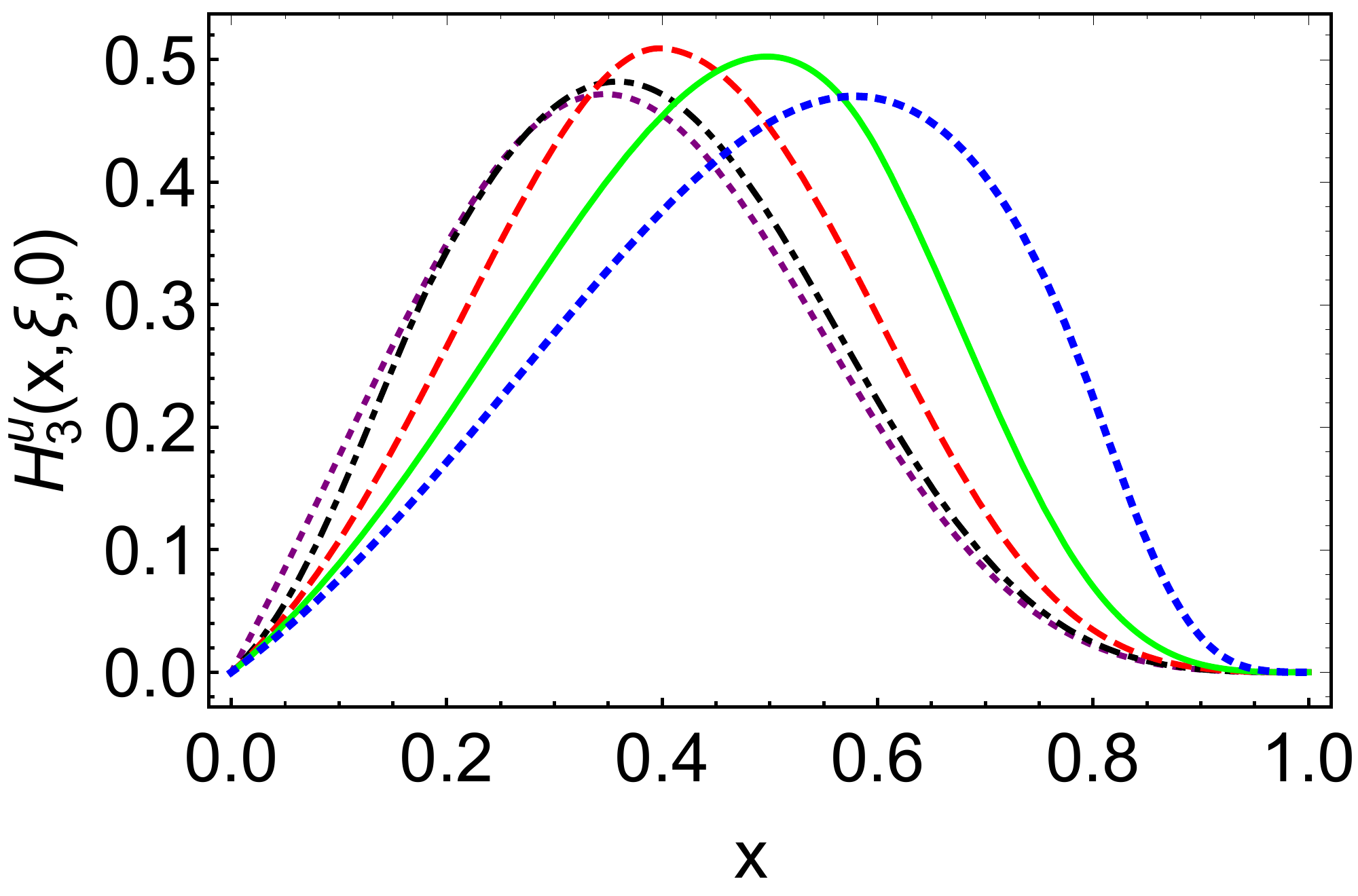}
\qquad
\includegraphics[width=0.47\textwidth]{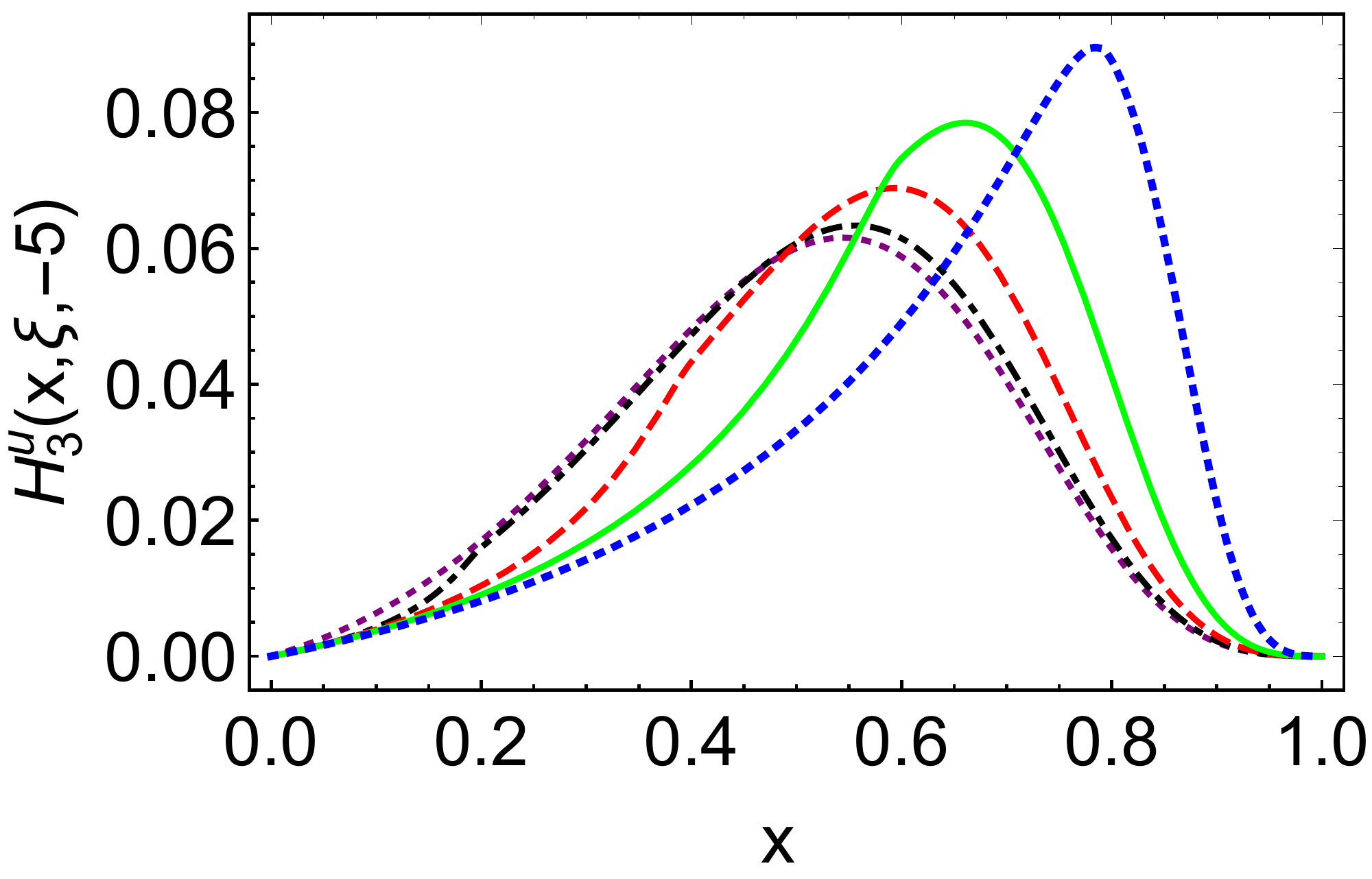}
\caption{$H_3^u(x,\xi,t)$ at $t=0$ GeV$^2$ (left panel) and $t=-5$ GeV$^2$ (right panel). $\xi=0$ (dotted purple), 0.2 (dot-dashed black),
 0.4 (dashed red), 0.6 (solid green), 0.8 (thick-dotted blue). }\label{h13} 
\end{figure*}

\begin{figure*}
\centering
\includegraphics[width=0.47\textwidth]{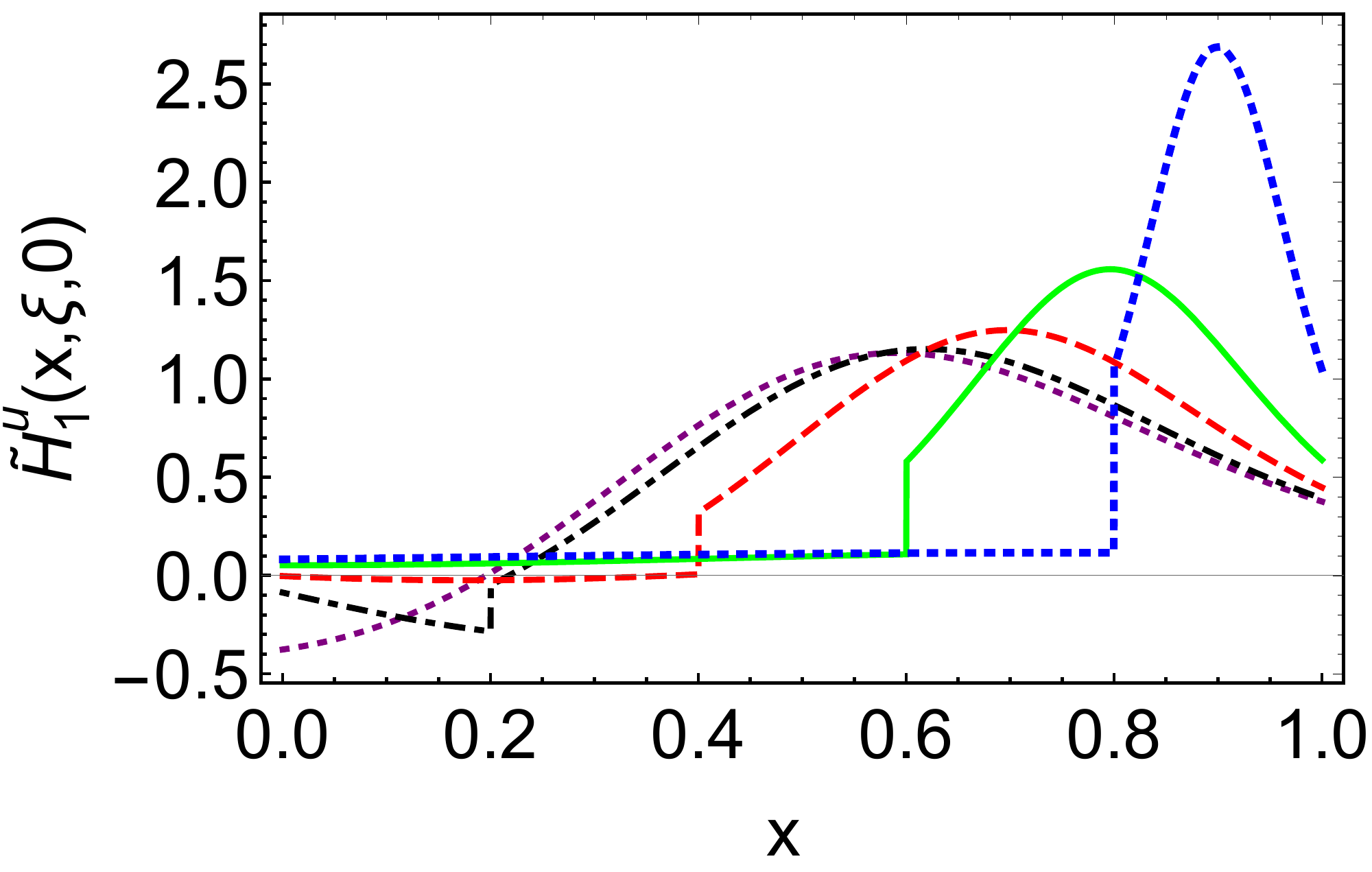}
\qquad
\includegraphics[width=0.47\textwidth]{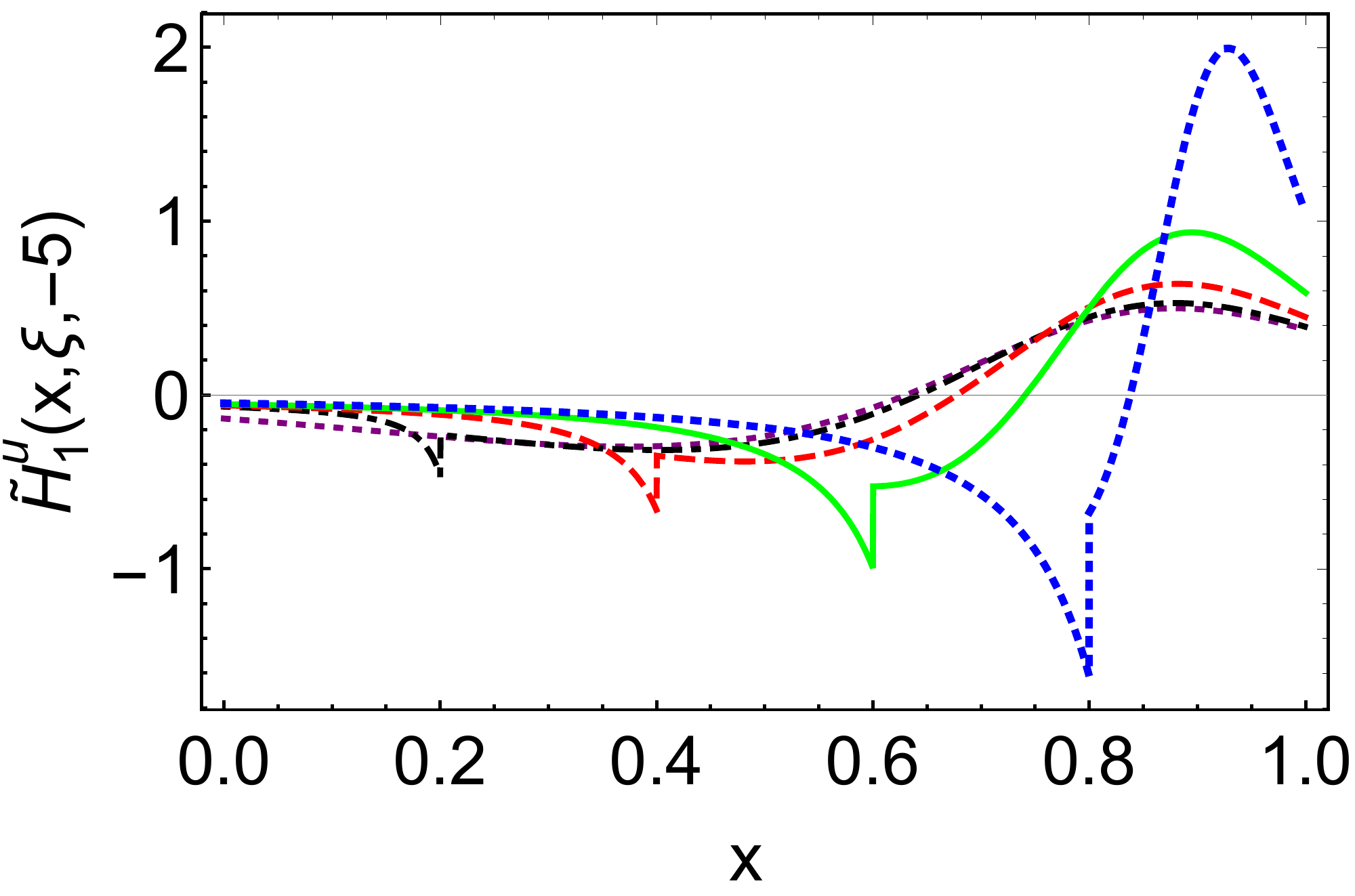}
\caption{$\tilde{H}_1^u(x,\xi,t)$ at $t=0$ GeV$^2$ (left panel) and $t=-5$ GeV$^2$ (right panel). $\xi=0$ (dotted purple), 0.2 (dot-dashed black),
 0.4 (dashed red), 0.6 (solid green), 0.8 (thick-dotted blue). }\label{th11} 
\end{figure*}

\begin{figure*}
\centering
\includegraphics[width=0.47\textwidth]{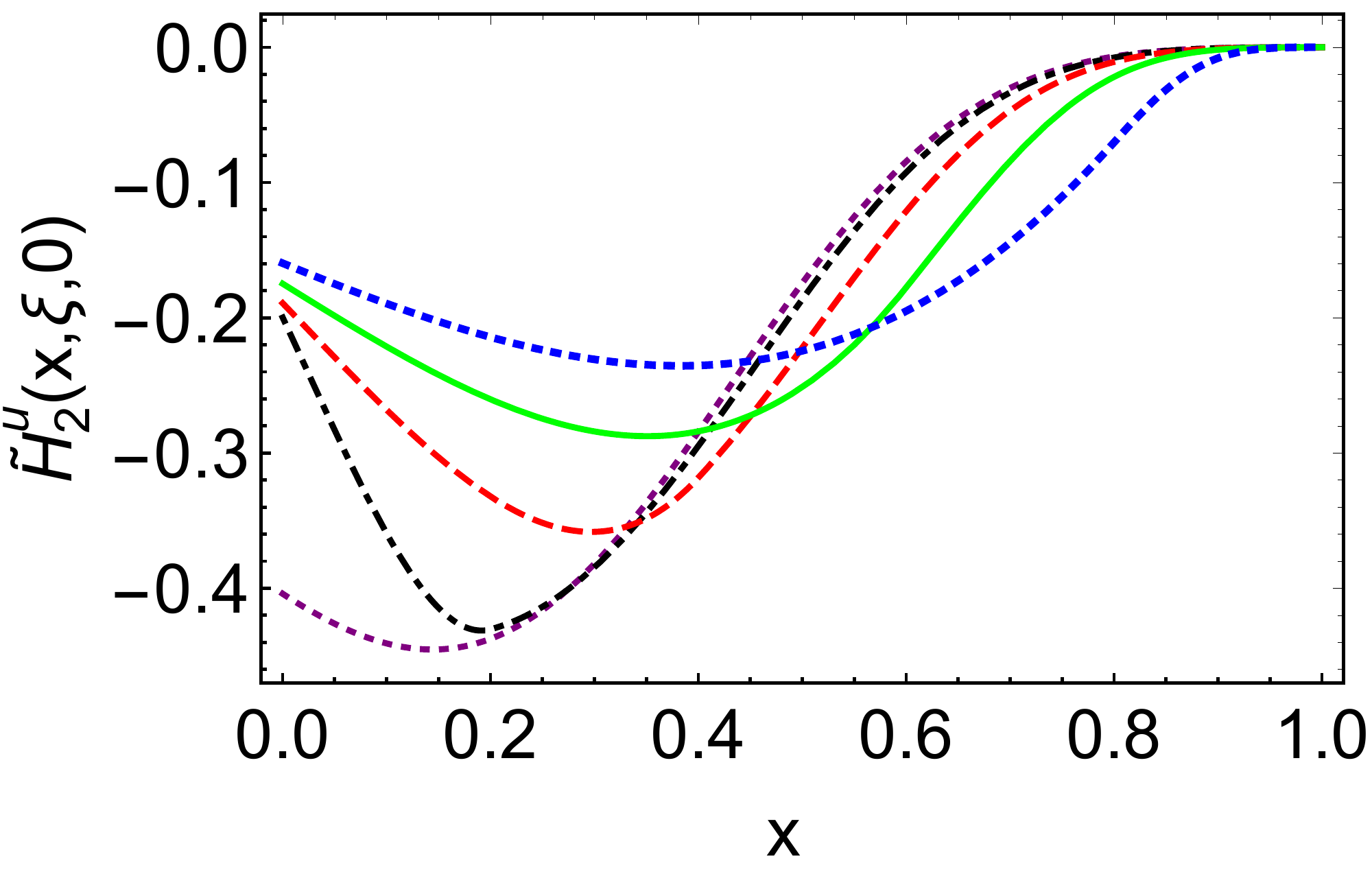}
\qquad
\includegraphics[width=0.47\textwidth]{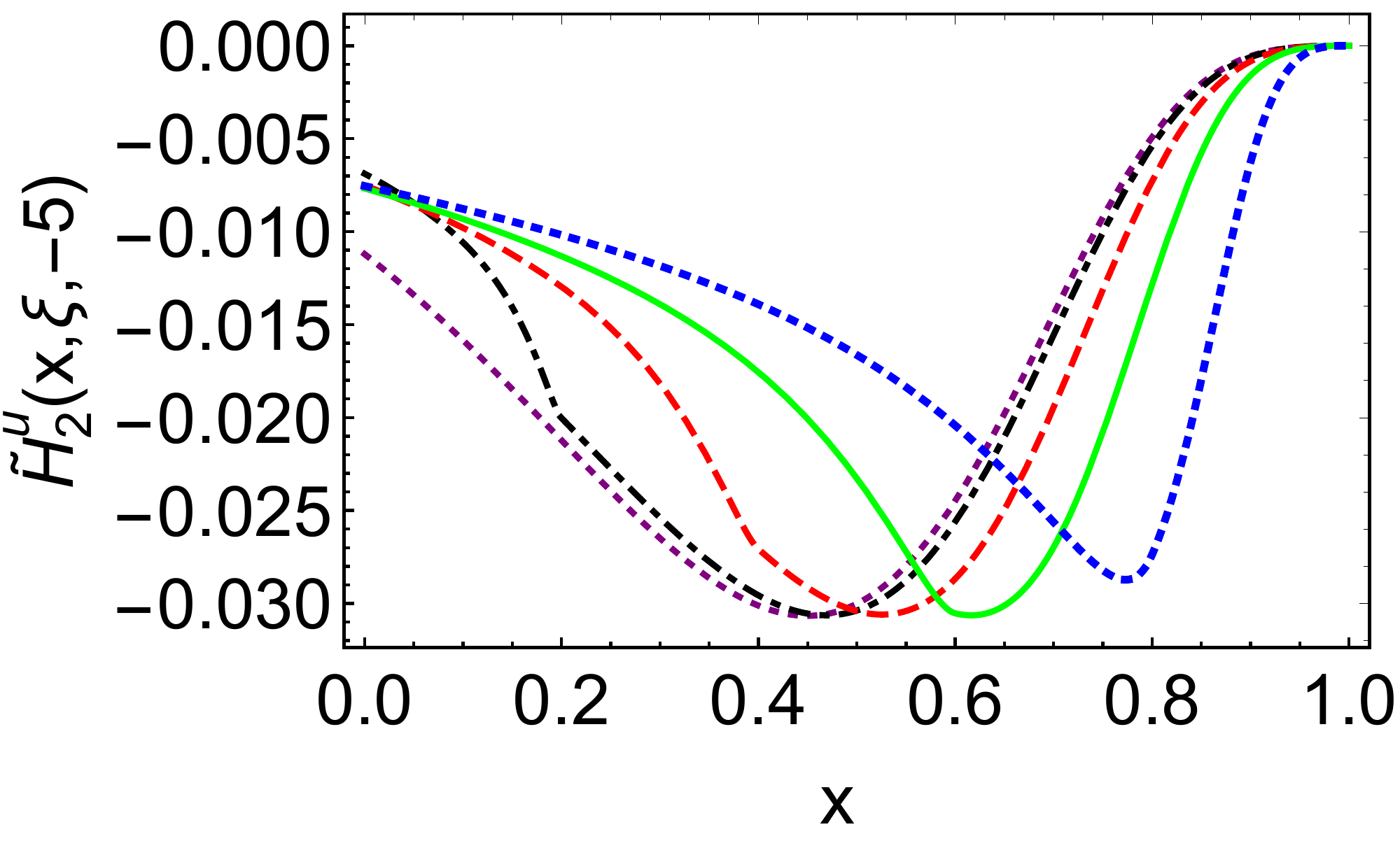}
\caption{$\tilde{H}_2^u(x,\xi,t)$ at $t=0$ GeV$^2$ (left panel) and $t=-5$ GeV$^2$ (right panel). $\xi=0$ (dotted purple), 0.2 (dot-dashed black),
 0.4 (dashed red), 0.6 (solid green), 0.8 (thick-dotted blue). }\label{th12} 
\end{figure*}

The pictures of $u$ quark GPDs are shown in Figs. \ref{h1}--\ref{th12}. From Figs.~\ref{h1}--\ref{th2} we can see that, in the region
$x\in[-1,-\xi]$, the unpolarized $H_i^u$ are zeroes, they have large values in the region $x\in[\xi,1]$, and they are small in the remaining region $x\in[-\xi,\xi]$, but not zeroes. The axial vector GPDs $\tilde{H}_i^u$ have similar behavior. Figs. \ref{h11}--\ref{th12} display GPDs at $t=0$ GeV$^2$ and $t=-5$ GeV$^2$ with $\xi=0, 0.2,0.4,0.8$, one can observe that $H_2^u$ and $\tilde{H}_1^u$ are not continuous at $x = \xi$, while
other GPDs are continuous at these points. Pion GPDs in the NJL model using Pauli-Villars regularization also show discontinuity at point $x=\xi$~\cite{Theussl:2002xp}, and even use other regularization schemes, the discontinuity of pion GPDs do not change~\cite{zhang2022regularization}. Ref.~\cite{Broniowski:2007si} calculated pion GPDs in both Spectral Quark Model and NJL model, the results also showed discontinuity.


\section{The properties of GPDs}\label{good}

\subsection{Symmetry properties}\label{qqS}
The unpolarized GPDs are neither even nor odd functions of $x$, the combinations,
\begin{subequations}\label{uconjugation}
\begin{align}
H_i^{I=0}(x,\xi,t)&=H_i^q(x,\xi,t)-H_i^q(-x,\xi,t)\,, \\
H_i^{I=1}(x,\xi,t)&=H_i^q(x,\xi,t)+H_i^q(-x,\xi,t),
\end{align}
\end{subequations}
correspond to the isoscalar (isovector) GPDs. The first combination has the charge conjugation parity $C=+1$, and the second has C-parity $C=-1$.

For the polarized quark GPDs, the combinations become
\begin{subequations}\label{conjugation}
\begin{align}
\tilde{H}_i^{I=0}(x,\xi,t)&=\tilde{H}_i^q(x,\xi,t)+\tilde{H}_i^q(-x,\xi,t)\,, \\
\tilde{H}_i^{I=1}(x,\xi,t)&=\tilde{H}_i^q(x,\xi,t)-\tilde{H}_i^q(-x,\xi,t).
\end{align}
\end{subequations}

Owing to the isospin symmetry and charge conjugation symmetry, quark GPDs are related by
\begin{subequations}\label{isospins}
\begin{align}
H_{i,\rho^+}^{u}(x,\xi,t)&=-H_{i,\rho^+}^{\bar{d}}(-x,\xi,t)\,, \\
\tilde{H}_{i,\rho^+}^{u}(x,\xi,t)&=\tilde{H}_{i,\rho^+}^{\bar{d}}(-x,\xi,t).
\end{align}
\end{subequations}
Now we study the even (odd) behaviors of GPDs with respect to the skewness $\xi$, or the behaviors of GPDs under the time reversal.
The behaviors are not uniform as pion GPDs in Refs.~\cite{Zhang:2020ecj,Zhang:2021mtn}, they are 
\begin{align}\label{as}
H_i(x,\xi,t)&=H_i(x,-\xi,t), \quad \quad (i=1,2,3,4),\nonumber\\
H_5(x,\xi,t)&=-H_5(x,-\xi,t),\nonumber\\
\tilde {H}_i(x,\xi,t)&=\tilde H_i(x,-\xi,t), \quad \quad (i=1,2,4),\nonumber\\
\tilde {H}_3(x,\xi,t)&=-\tilde {H}_3(x,-\xi,t).
\end{align}
It is worth noting that time reversal invariance determines the behaviors of GPDs under sign change of $\xi$ and fixes the phase of GPDs
but does not restrict the number of GPDs in the nonforward case~\cite{Diehl:2001pm}.

\subsection{Forward limit}\label{qqQ}
When the initial and final momentum equal, $p=p^{\prime}$, $\xi=0$, $t=0$, GPDs should reduce to usual PDFs. The relationships
$\varepsilon\cdot p=\varepsilon^{\prime} \cdot p^{\prime}=0$ in the forward limit become $\varepsilon\cdot P=\varepsilon^{\prime} \cdot P=0$.
From Eqs. (\ref{aX1}) and (\ref{aX2}) one can see that for the unpolarized GPDs only $H_1$ and $H_5$ survive, for the polarized GPDs,
only $\tilde{H}_1$ is preserved. In the leading order, there are three independent distribution functions in DIS, $F_1(x)$, $b_1(x)$, $g_1(x)$.
The probabilistic interpretation in terms of quark densities has the form
\begin{subequations}\label{fl}
\begin{align}
F_1^u(x)&=\frac{q^1(x)+q^{-1}(x)+q^0(x)}{3}\,, \\
b_1^u(x)&=q^0(x)-\frac{q^1(x)+q^{-1}(x)}{2}\,, \\
g_1^u(x)&=q_{\uparrow}^1(x)-q_{\uparrow}^{-1}(x),
\end{align}
\end{subequations}
for $x>0$. $q^{\lambda}(x)$ stands for unpolarized PDF, which is defined as $q^{\lambda}=q_{\uparrow}^{\lambda}+q_{\downarrow}^{\lambda}$,
where $\uparrow$ ($\downarrow$) represents up (down) spin projection along the moving direction of $\rho$ meson in the infinite momentum frame.
For $x<0$, the previous two above equations with an overall sign change give the antiquark distributions at $-x$, the last equation stands for the antiquark distribution at $-x$.

The relationships between GPDs and PDFs using the results for helicity amplitude in the forward limit give
\begin{subequations}\label{pc}
\begin{align}\label{as}
F_1^u(x)&=\frac{1}{2}H_1^u(x,0,0)\,, \\
b_1^u(x)&=H_5^u(x,0,0)\,, \\
g_1^u(x)&=\frac{1}{2}\tilde{H}_1^u(x,0,0).
\end{align}
\end{subequations}
In the leading order, the single flavor DIS distribution function $F_1^u(x)$ is $1/2$ of the probability to find a quark with momentum
fraction $x$ and follows the Callan-Gross relation~\cite{Berger:2001zb,Mondal:2017lph}. The single flavor distribution function $b_1^u(x)$,
which measures the difference in the spin projection of the $\rho$ meson, depends solely on the quark-spin-averaged distribution.

In the present approach, we obtain the three distribution functions as follows,
\begin{align}\label{dis1}
F_1^u(x)&=\frac{3Z_{\rho}}{8\pi^2}(1-2x+2x^2)\bar{\mathcal{C}}_1(\sigma_6) \nonumber\\
&+\frac{3Z_{\rho}}{4\pi^2} x(1-x) m_{\rho}^2\frac{1}{\sigma_6}\bar{\mathcal{C}}_2(\sigma_6) ,
\end{align}
\begin{align}\label{dis2}
b_1^u(x)
&=-\frac{3Z_{\rho}}{4\pi^2}(1-6x+6x^2)\bar{\mathcal{C}}_1(\sigma_6) \nonumber\\
&-\frac{3Z_{\rho}}{2\pi^2}x(1-x)(1-2x)^2 m_{\rho}^2  \frac{1}{\sigma_6}\bar{\mathcal{C}}_2(\sigma_6) ,
\end{align}
\begin{align}\label{dis3}
g_1^u(x)&=\frac{3Z_{\rho }}{8\pi ^2}(2x-1)\bar{\mathcal{C}}_1(\sigma_6)\nonumber\\
&+\frac{3 Z_{\rho }}{4\pi^2}(1-x)x\,m_{\rho}^2 \frac{\bar{\mathcal{C}}_2(\sigma_6)}{\sigma_6}.
\end{align}
Based on the above expressions, it is easy to obtain the sum rules for $H_1^u$ and $H_5^u$
\begin{subequations}\label{spdf}
\begin{align}\label{as}
\int_{-1}^1H_1^u(x,0,0)\mathrm{d}x =1\,, \\
\int_{-1}^1H_5^u(x,0,0)\mathrm{d}x =0,
\end{align}
\end{subequations}
Under the assumption that the quark sea does not contribute to the integration $\int_0^1 b_1(x)\mathrm{d}x$,
one can obtain the parton model sum rule,
\begin{align}\label{as}
\int_0^1 b_1(x)\mathrm{d}x=0.
\end{align}
Using the conventional definitions of distribution functions, we obtain
\begin{subequations}\label{cpdf}
\begin{align}
& F_1(x)=\sum_q e_q^2F_1^q(x)\,, \\
& b_1(x)=\frac{1}{2}\sum_q e_q^2b_1^q(x)\,, \\
& g_1(x)=\sum_q e_q^2g_1^q(x).
\end{align}
\end{subequations}
The polarized parton function $g_1(x)$ gives the fraction of spin carried by quarks, follows the relationship
\begin{align}\label{as}
g_1(x)=\sum_q e_q^2 g_1^q(x)=\frac{1}{2}(e_u^2+e_{\bar{d}}^2)\Delta u(x),
\end{align}
\begin{align}\label{fs}
\Delta q\equiv \int_0^1(g_1^u(x)+g_1^{\bar{d}}(x)) \mathrm{d}x =\int_0^1\Delta u(x) \mathrm{d}x,
\end{align}
we obtain $\Delta q=0.551$, which describes the fraction of spin carried by the constituent quark and antiquark in $\rho$ meson is not the expected value 1. Here, we suggest that the orbital angular momentum of partons maybe also important in making up the spin
of $\rho$ meson.

\begin{figure}
\centering
\includegraphics[width=0.47\textwidth]{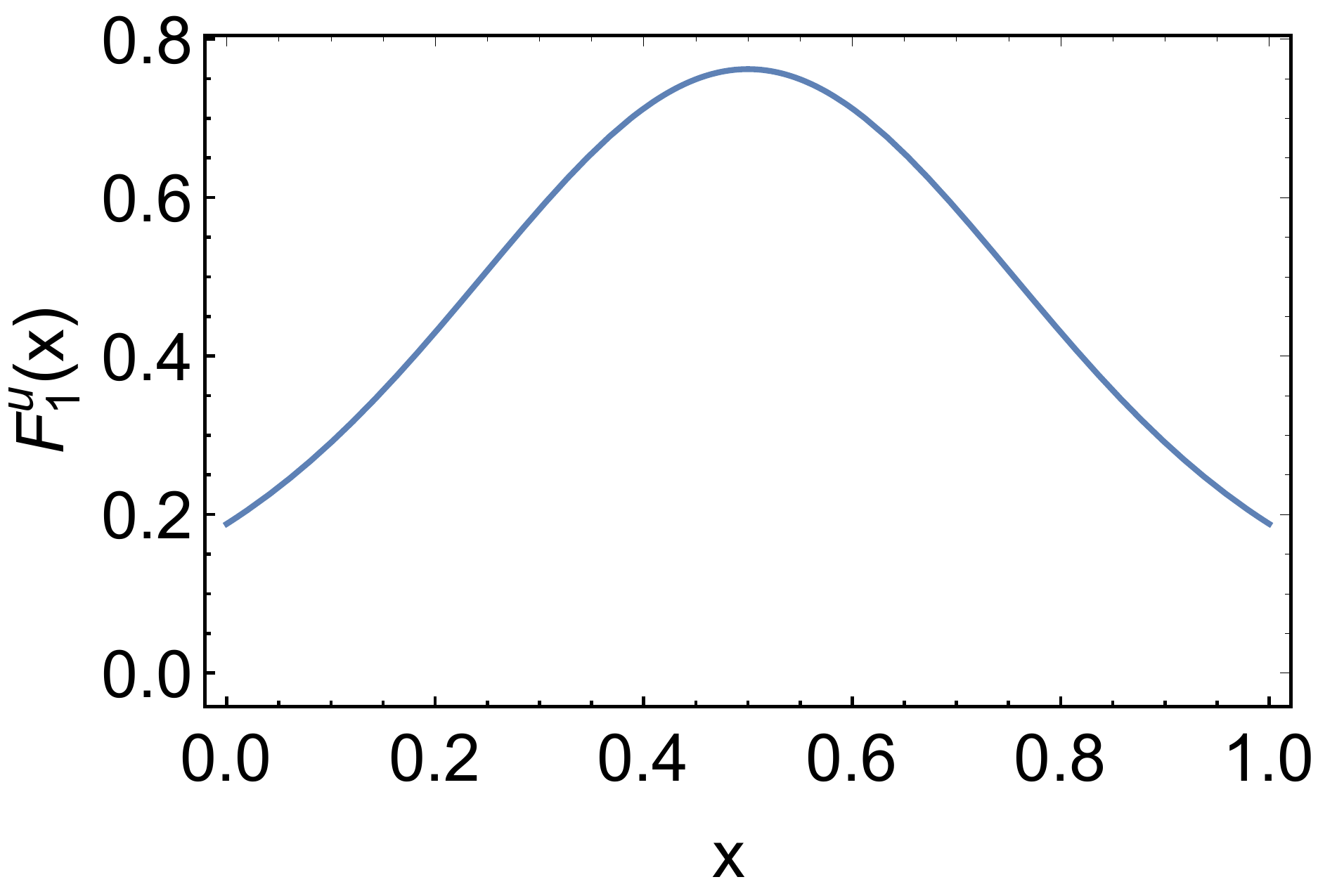}
\caption{The $u$ quark DIS distribution function $F_1^u(x)$.}\label{disf1}
\end{figure}

\begin{figure}
\centering
\includegraphics[width=0.47\textwidth]{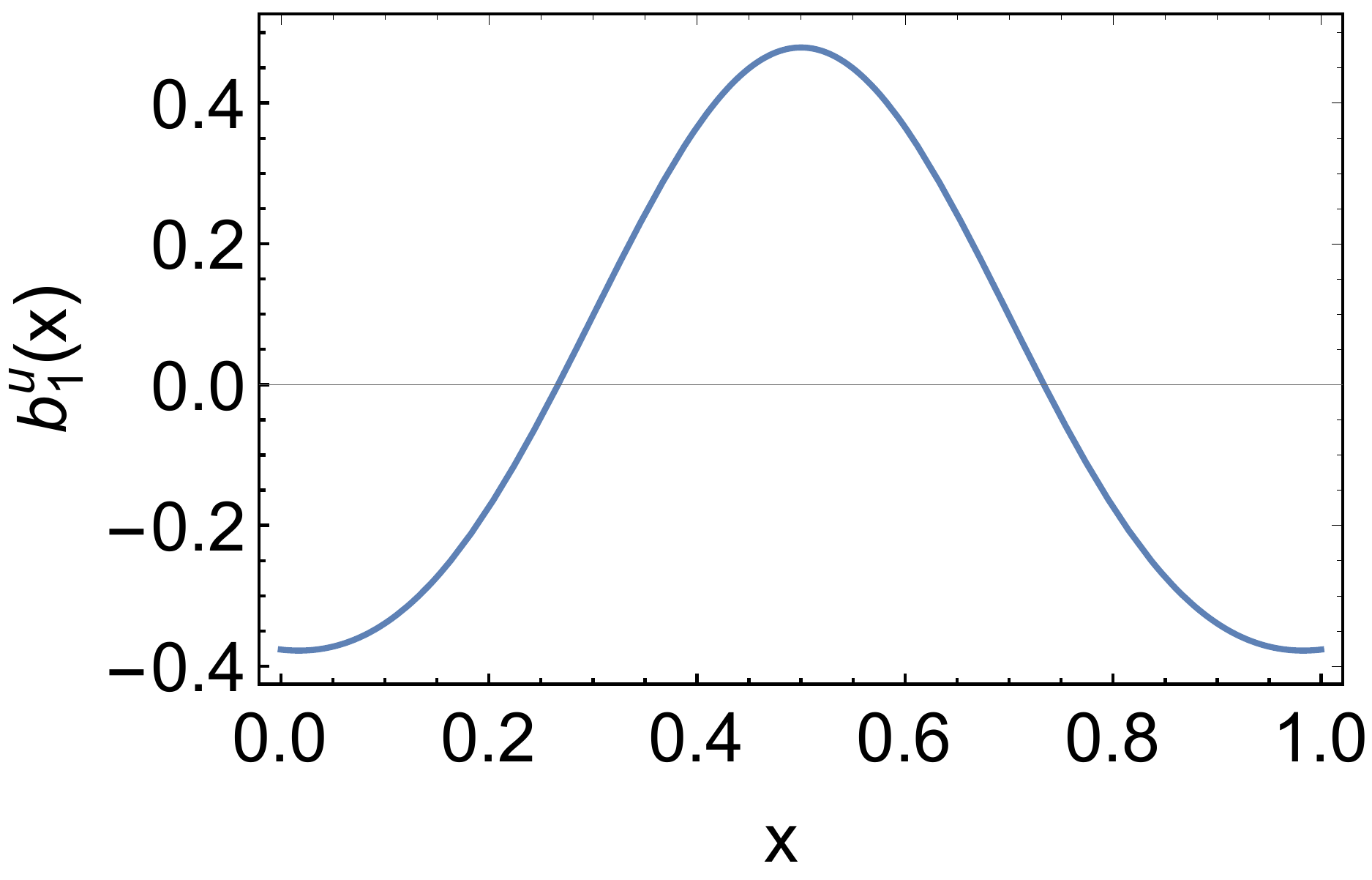}
\caption{The $u$ quark DIS distribution function $b_1^u(x)$.}\label{disb1}
\end{figure}

\begin{figure}
\centering
\includegraphics[width=0.47\textwidth]{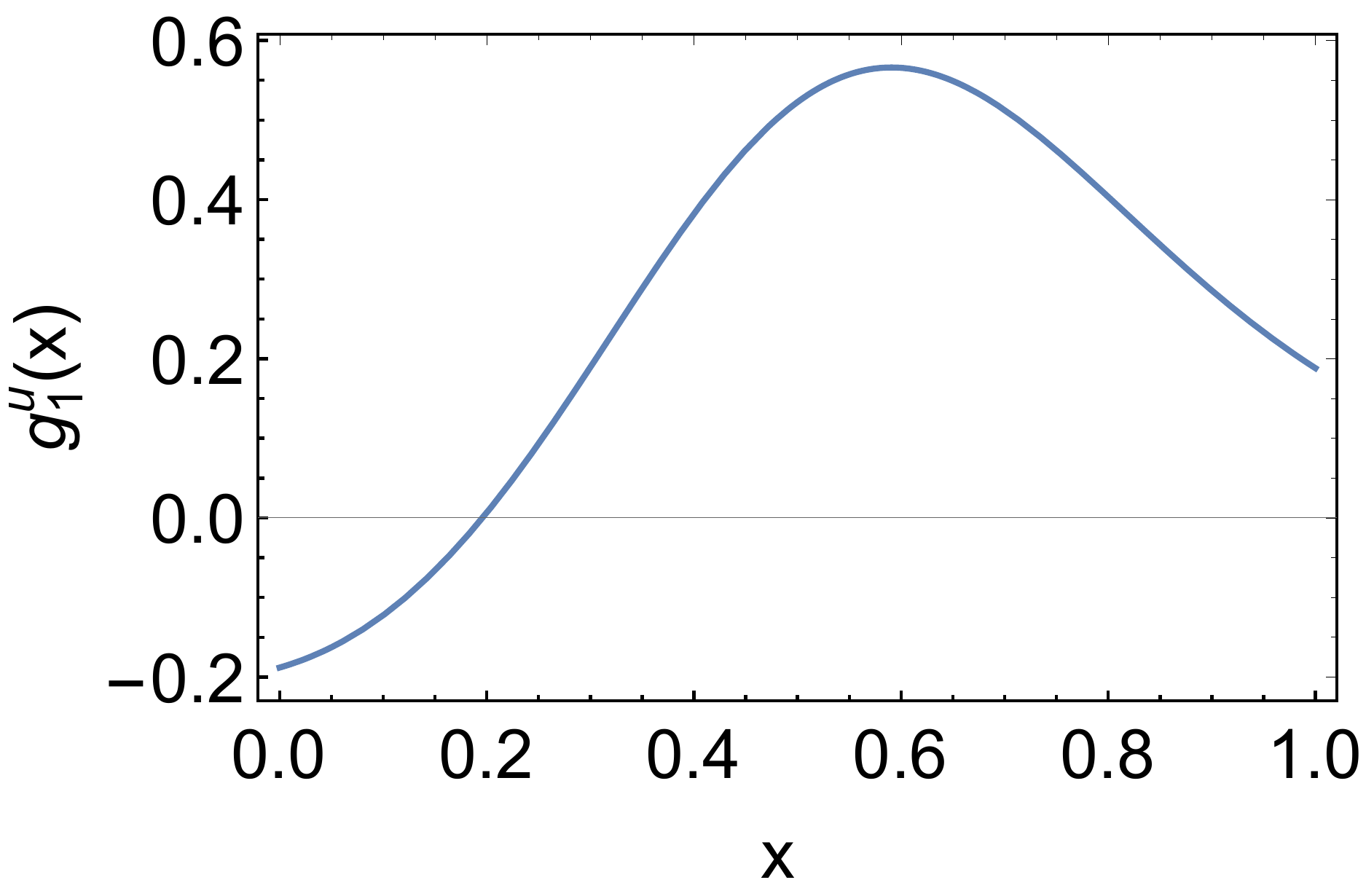}
\caption{The $u$ quark DIS distribution function $g_1^u(x)$.}\label{updf}
\end{figure}

We have illustrated $F_1^u(x)$, $b_1^u(x)$ and $g_1^u(x)$ in Figs. \ref{disf1}--\ref{updf}. Our results show that $F_1^u(x)$ is not zero at $x=0$, which can also be seen from the GPDs diagrams of $H_1^u$. The reason may be that the gluon degrees of freedom have been frozen into effective point-like interactions between quarks in the NJL model. We can see that $F_1^u(x)$ is approximately symmetric with respect to $x \sim 1/2$.

\subsection{Form factors}\label{Aqq}
The definitions of vector and axial currents are
\begin{widetext}
\begin{subequations}\label{pc}
\begin{align}\label{bF3}
\langle p^{\prime},\lambda^{\prime}|\bar{\psi}(0)\gamma^{\alpha}\psi(0)|p,\lambda\rangle&=-(\epsilon^{\prime *} \cdot \epsilon)2 P^{\alpha}F_1(t)+[\epsilon^{\alpha}( \epsilon^{\prime *}\cdot 2P)+\epsilon^{\prime *\alpha}( \epsilon\cdot 2P)]F_2(t)\nonumber\\
&-(\varepsilon\cdot 2P )(\epsilon^{\prime *}\cdot 2P)\frac{P^{\alpha}}{m_{\rho}^2}F_3(t)\,, \\
\langle p^{\prime},\lambda^{\prime}|\bar{\psi}(0)\gamma^{\alpha}\gamma^5\psi(0)|p,\lambda\rangle&=-2i\epsilon_{\alpha  \nu\mu \beta }\epsilon^{\prime *\nu} \epsilon^{\mu}P^{\beta } \tilde{F}_1(t)+4i\epsilon_{\alpha  \nu\mu \beta }\Delta^{\nu} P^{\mu} \frac{\epsilon^{\beta }( \epsilon^{\prime *}\cdot P)+(\epsilon\cdot P) \epsilon^{\prime *\beta}}{m_{\rho}^2}\tilde{F}_2(t).
\end{align}
\end{subequations}
\end{widetext}
For $\rho$ meson, the general forms of vector and axial currents can be written as
\begin{align}\label{bF3}
j_{\rho,v}^{\alpha,\mu\nu }&=[-g^{\mu\nu }F_1(t)-\frac{2P^{\mu}P^{\nu }}{m_{\rho}^2}F_3(t)](p^{\alpha}+p^{\prime \alpha})\nonumber\\
&+2(P^{\nu}g^{\alpha \mu}+P^{\mu}g^{\alpha\nu})F_2(t),
\end{align}
\begin{align}\label{bF3}
j_{\rho,a}^{\alpha,\mu\nu }&=-2i \epsilon_{\alpha \mu\nu \rho }P^{\rho} \tilde{F}_1(t)\nonumber\\
&+4i\Delta^{\rho}P^{\beta}\frac{(\epsilon_{\alpha \rho\beta \nu}P^{\mu}+\epsilon_{\alpha \rho\beta \mu}P^{\nu})}{m_{\rho}^2}\tilde{F}_2(t),
\end{align}
where the Lorentz indices $\alpha,\mu,\nu$ represent the polarizations of photon, initial and final $\rho$ meson, separately. The form factors in the above expressions can be obtained from the integration of GPDs over the momentum fraction $x$,
\begin{subequations}\label{bF31}
\begin{align}
\int_{-1}^1\mathrm{d}x H_i^q(x,\xi,t)&=F_i^q(t), \quad \quad (i=1,2,3) \,, \\
\int_{-1}^1\mathrm{d}x \tilde H_i^q(x,\xi,t)&=\tilde {F}_i^q(t), \quad \quad (i=1,2)\,, \\
\int_{-1}^1\mathrm{d}x H_i^q(x,\xi,t)&=0, \quad \quad (i=4,5)\,, \\
\int_{-1}^1\mathrm{d}x \tilde{H}_i^q(x,\xi,t)&=0, \quad \quad (i=3,4),
\end{align}
\end{subequations}
our GPDs in Eqs. (\ref{ugpd1})-(\ref{pgpd2}) satisfy the sum rules, in addition, they also satisfy the polynomiality condition, which is related to
the gravitational form factors and higher Mellin moments of GPDs.

The conventional FFs are obtained through weighting the electromagnetic charges and then summing over flavors:
\begin{align}
F_i(t)
&=e_u \int_{-1}^1 \mathrm{d}x H_i^u(x,\xi,t)+e_{d}\int_{-1}^1 \mathrm{d}x H_i^d(x,\xi,t)\nonumber\\
&= \int_{-1}^1 \mathrm{d}x H_i^{I=1}(x,\xi,t).
\end{align}
From Eqs. (\ref{ugpd1})-(\ref{pgpd2}) we obtain $F_i^u(t)$
\begin{align}\label{bF3}
F_{1 }^u(t)&=\frac{N_cZ_{\rho }}{4\pi^2}\int_0^1 \mathrm{d}x \,\bar{\mathcal{C}}_1(\sigma_6) \nonumber\\
&+ \frac{N_cZ_{\rho }}{2\pi^2}\int_0^1 \mathrm{d}x\int_0^{1-x} \mathrm{d}y\,\frac{1}{\sigma_7}\bar{\mathcal{C}}_2(\sigma_7) \nonumber\\
&\times ((1-x-y)m_{\rho}^2+\frac{(x+y)}{2}t)\nonumber\\
&-\frac{N_cZ_{\rho }}{2\pi^2}\int_0^1 \mathrm{d}x\int_0^{1-x} \mathrm{d}y\,(1-x-y)\,\bar{\mathcal{C}}_1(\sigma_7),
\end{align}
\begin{align}\label{bF3}
F_{2}^u(t)&= \frac{N_cZ_{\rho } }{4 \pi^2} \int_0^1 \mathrm{d}x\, \bar{\mathcal{C}}_1(\sigma_1)\nonumber\\
&+\frac{N_cZ_{\rho } }{4 \pi^2} \int_0^1 \mathrm{d}x \,\bar{\mathcal{C}}_1(\sigma_6) \nonumber\\
&+\frac{N_cZ_{\rho }}{2\pi^2} \int_0^1 \mathrm{d}x\int_0^{1-x} \mathrm{d}y \,(m_{\rho}^2+\frac{(x+y)}{2}t)  \frac{\bar{\mathcal{C}}_2(\sigma_7)}{\sigma_7} \nonumber\\
&-\frac{N_cZ_{\rho }}{2\pi^2} \int_0^1 \mathrm{d}x\int_0^{1-x} \mathrm{d}y\, (x+y) \, \bar{\mathcal{C}}_1(\sigma_7),
\end{align}
\begin{align}\label{bF3}
F_{3}^u(t)&=\frac{4 m_{\rho}^2N_cZ_{\rho }}{\pi^2} \int_0^1 \mathrm{d}x\int_0^{1-x} \mathrm{d}y \nonumber\\
&\times   x\,y\,(1-x-y)\frac{1}{\sigma_7}\bar{\mathcal{C}}_2(\sigma_7),
\end{align}
\begin{align}\label{bF3}
\tilde{F}_{1 }^u(t)&=\frac{N_cZ_{\rho }}{4\pi^2}\int_0^1 \mathrm{d}x\, \bar{\mathcal{C}}_1(\sigma_6) \nonumber\\
&-\frac{N_cZ_{\rho }}{2\pi^2} \int_0^1 \mathrm{d}x \int_0^{1-x} \mathrm{d}y\,  \bar{\mathcal{C}}_1(\sigma_7)\nonumber\\
&-\frac{N_cZ_{\rho }}{4\pi^2} \int_0^1 \mathrm{d}x \int_0^{1-x} \mathrm{d}y\, \frac{\bar{\mathcal{C}}_2(\sigma_7)}{\sigma_7} \nonumber\\
&\times (2(x+y-1)\,m_{\rho}^2-(x+y)\,t),
\end{align}
\begin{align}\label{bF3}
\tilde{F}_{2 }^u(t)&=-\frac{N_cZ_{\rho }}{\pi^2}\int_0^1 \mathrm{d}x \int_0^{1-x} \mathrm{d}y \, x\,y \,m_{\rho}^2 \frac{\bar{\mathcal{C}}_2(\sigma_7)}{\sigma_7}.
\end{align}
Due to the isospin symmetry, $\tilde{F}_{i }^u=\tilde{F}_{i }^{\bar{d}}$ in $\rho$ meson, the contributions of $u$ and $\bar{d}$ quarks to the general
axial vector FFs $\tilde{F}_{i }^{I=1} =\tilde{F}_{i }^u-\tilde{F}_{i }^{\bar{d}}$ cancel each other, $\tilde{F}_{i }^{I=1} =0$, but
$\tilde{F}_{i }^{I=0}=\tilde{F}_{i }^u+\tilde{F}_{i }^{\bar{d}}=2\tilde{F}_{i }^u$.

The Sachs-like charge, magnetic, and quadruple form factors for $\rho$ meson are given by
\begin{align}\label{sff}
G_C(t)&=F_1(t)+\frac{2}{3}\eta \,G_Q(t),\nonumber\\
G_M(t)&=F_2(t),\nonumber\\
G_Q(t)&=F_1(t)+(1+\eta)F_3(t)-F_2(t),
\end{align}
where $\eta=-t/(4m_{\rho}^2)$.

In Ref.~\cite{Brodsky:1992px}, they derived a relation for form factors of spin-$1$ particles at the large $Q^2$, which is, at large timelike or
spacelike momenta, the ratio of form factors for $\rho$ meson should behave as
\begin{align}\label{lsff}
G_C(t):G_M(t):G_Q(t)=(1+\frac{2}{3}\eta):2:-1
\end{align}
where corrections are of the orders $\Lambda_{\text{QCD}}/Q$ and $\Lambda_{\text{QCD}}/m_{\rho}$. In the present calculation, at $Q^2=10$ GeV$^2$,
$1+\frac{2}{3}\eta=-1.811$, $G_C/G_M=-0.888$, $G_C/G_Q=2.244$, $G_M/G_Q=-2.527$. These results are consistent with the relation.

When $Q^2=0$, these form factors give the charge, magnetic moment, and quadruple moment of $\rho$ meson.
\begin{align}\label{bF3}
G_C(0)=1, \quad G_M(0)= \mu_{\rho}, \quad G_Q(0)=\mathcal{Q}_{\rho},
\end{align}
where the charge is in units of the fundamental charge $e$, $\mu_{\rho}$ and $\mathcal{Q}_{\rho}$ are the $\rho$ meson magnetic dipole and quadrupole moments
in units of $e/2m_{\rho}$ and $e/m_{\rho}^2$, respectively.
The charge, magnetic and quadruple radius, $\langle r_C^2\rangle$, $\langle r_M^2\rangle$, $\langle r_Q^2\rangle$, are given by
\begin{align}\label{bF3}
\langle r^2\rangle=-\frac{6}{\kappa}\frac{\partial G(Q^2)}{\partial Q^2}|_{Q^2=0}
\end{align}
when $G(0)=0$, $\kappa=1$, if $G(0)\neq 0$, $\kappa=G(0)$. $G(Q^2)$ is one of the above form factors. Our results for the Sachs-like charge, magnetic moment, quadruple moment and the charge, magnetic, and quadruple radius of $\rho$ meson
are tabulated in Table \ref{tb2}.

For point-like spin-$1$ particles, the magnetic dipole and the electric quadrupole moments are $\mu = 2$ and $\mathcal{Q} = -1$.
Our results give $\mu_{\rho} = 2.083$, which is smaller than $\mu_{\rho} = 2.20$ in Ref.~\cite{Biernat:2014dea} but larger than
$\mu_{\rho} = 1.92$ in Ref.~\cite{Choi:2004ww}. Our quadrupole moment $\mathcal{Q}_{\rho}=-0.871$, which is about $13 \%$ percent larger than the canonical value $-1$ because of the relativistic effects. Our results are in accordance with that of Ref.~\cite{Cloet:2014rja}, where they calculated the $\rho$ meson Sachs-like FFs in the NJL model. The consistency means that the same FFs from GPDs are obtained in the two approaches.

\begin{figure}
\centering
\includegraphics[width=0.5\textwidth]{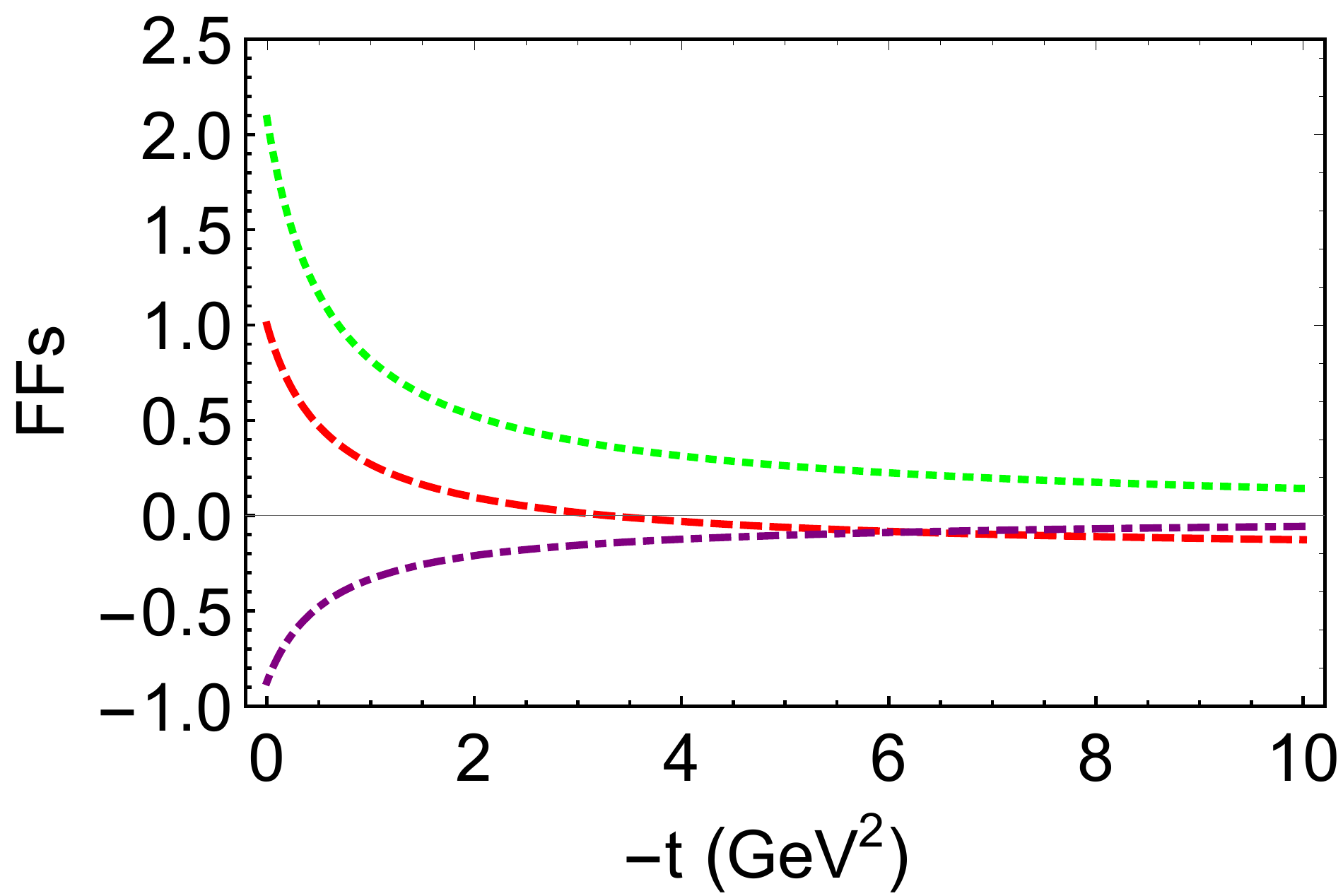}
\caption{The dressed Sachs-like charge (dashed red), magnetic (dotted green), and quadruple (dot-dashed purple) form factors.}\label{rhopl}
\end{figure}

\begin{figure}
\centering
\includegraphics[width=0.47\textwidth]{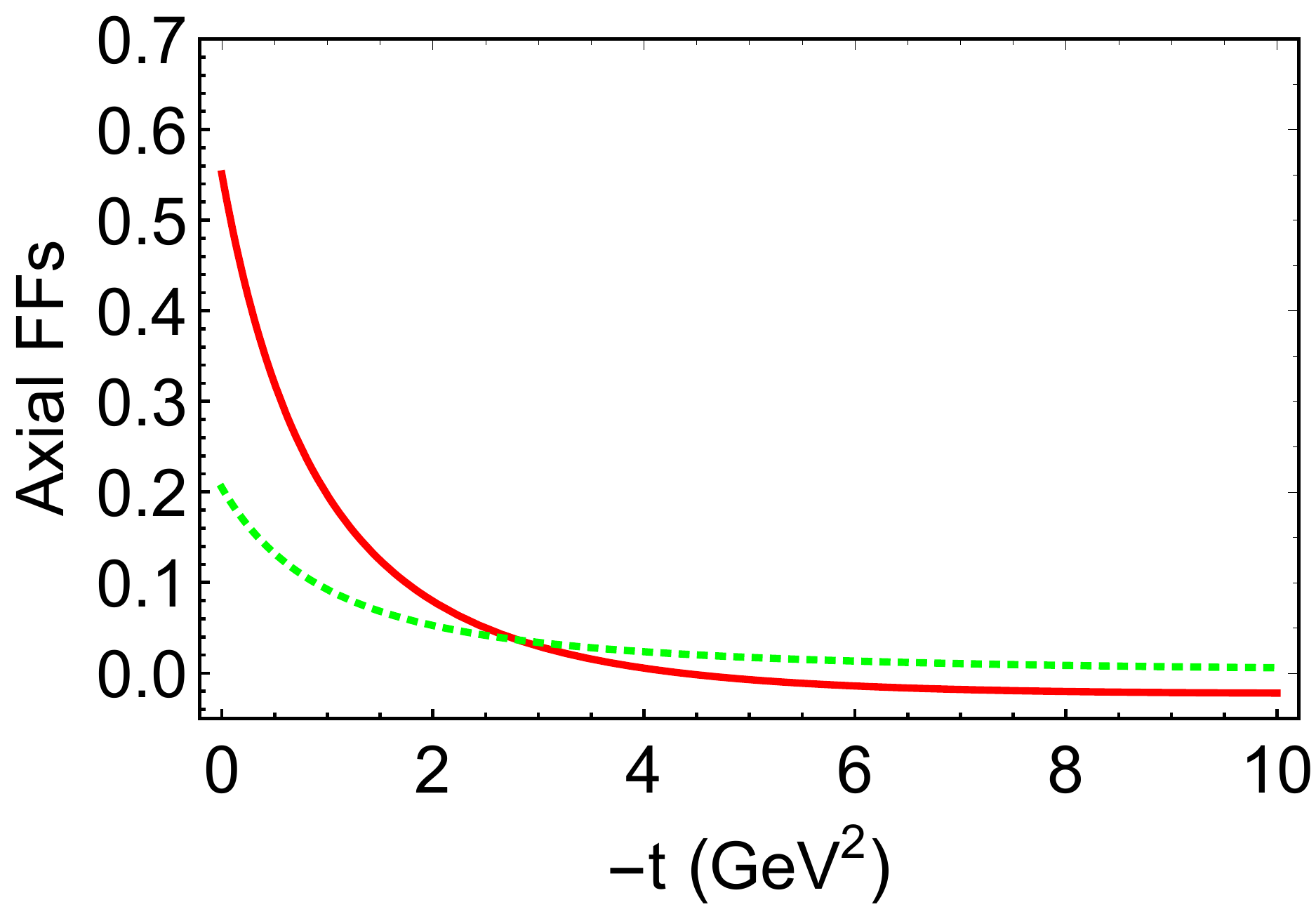}
\caption{The $u$ quark axial form factors, $\tilde{F}_1^u(t)$ (solid red), $-\tilde{F}_2^u(t)$ (dotted green), of $\rho$ meson.}\label{aff}
\end{figure}

In Fig.~\ref{rhopl}, we plot the diagrams of the three Sachs-like FFs. The dressed FFs are defined as
\begin{eqnarray*}
G_C^D(t) & = & G_C(t)F_{1\rho}(t), \\
G_M^D(t) & = & G_M(t)F_{1\rho}(t), \\
G_Q^D(t) & = & G_Q(t)F_{1\rho}(t),
\end{eqnarray*}
where $F_{1\rho}(t)$ is defined in Eq. (\ref{bsam}). The behaviors of FFs agree with the previous results, such as Refs.~\cite{Cardarelli:1994yq,Choi:2004ww,Cloet:2014rja,Biernat:2014dea,Sun:2017gtz}. In the present calculation,
the crossing point of the Sachs-like charge is around $t\approx-3.3$ GeV$^2$.
In Fig. \ref{aff}, we plot the diagrams of axial vector FFs $\tilde{F}_{1 }^u(t)$ and $\tilde{F}_{2 }^u(t)$, where the starting points are
$\tilde{F}_{1}^u(0)=0.551$ and $\tilde{F}_{2 }^u(0)=-0.205$. The polarized GPDs are also studied in Ref.~\cite{Sun:2018ldr}, they gave
$\tilde{F}_{1 }^u(0)=0.86$ and $\tilde{F}_{2 }^u(0)=-0.16$ in the light-front constituent quark model.
$\tilde{F}_{1 }^u(0)$ equals $\Delta q$ in Eq. (\ref{fs}), which means the fraction of spin carried by the constituent quark and antiquark of
$\rho$ meson in the NJL model is smaller than in the light-front constituent quark model.

\begin{center}
\begin{table}
\caption{Results for the magnetic moment, quadruple moment, and the charge, magnetic, and quadruple radius of the $\rho$ meson. The radius are in units of fm; the magnetic moment has units $e/(2m_{\rho})$ and the quadruple moment $e/m_{\rho}^2$. }\label{tb2}
\begin{tabular}{p{0.9cm} p{1.1cm} p{0.9cm} p{0.9cm}p{0.9cm}p{0.9cm}p{0.9cm}p{0.9cm}}
\hline\hline
$\mu$&$\mathcal{Q}_{\rho}$&$r_C$&$r_C^D$&$r_M$&$r_M^D$&$r_Q$&$r_Q^D$\\
\hline
2.082&$-$0.871&0.517&0.672&0.440&0.615&0.449&0.621\\
\hline\hline
\end{tabular}
\end{table}
\end{center}

\subsection{Impact parameter dependent PDFs}\label{Aqq}
\begin{figure}
\centering
\includegraphics[width=0.47\textwidth]{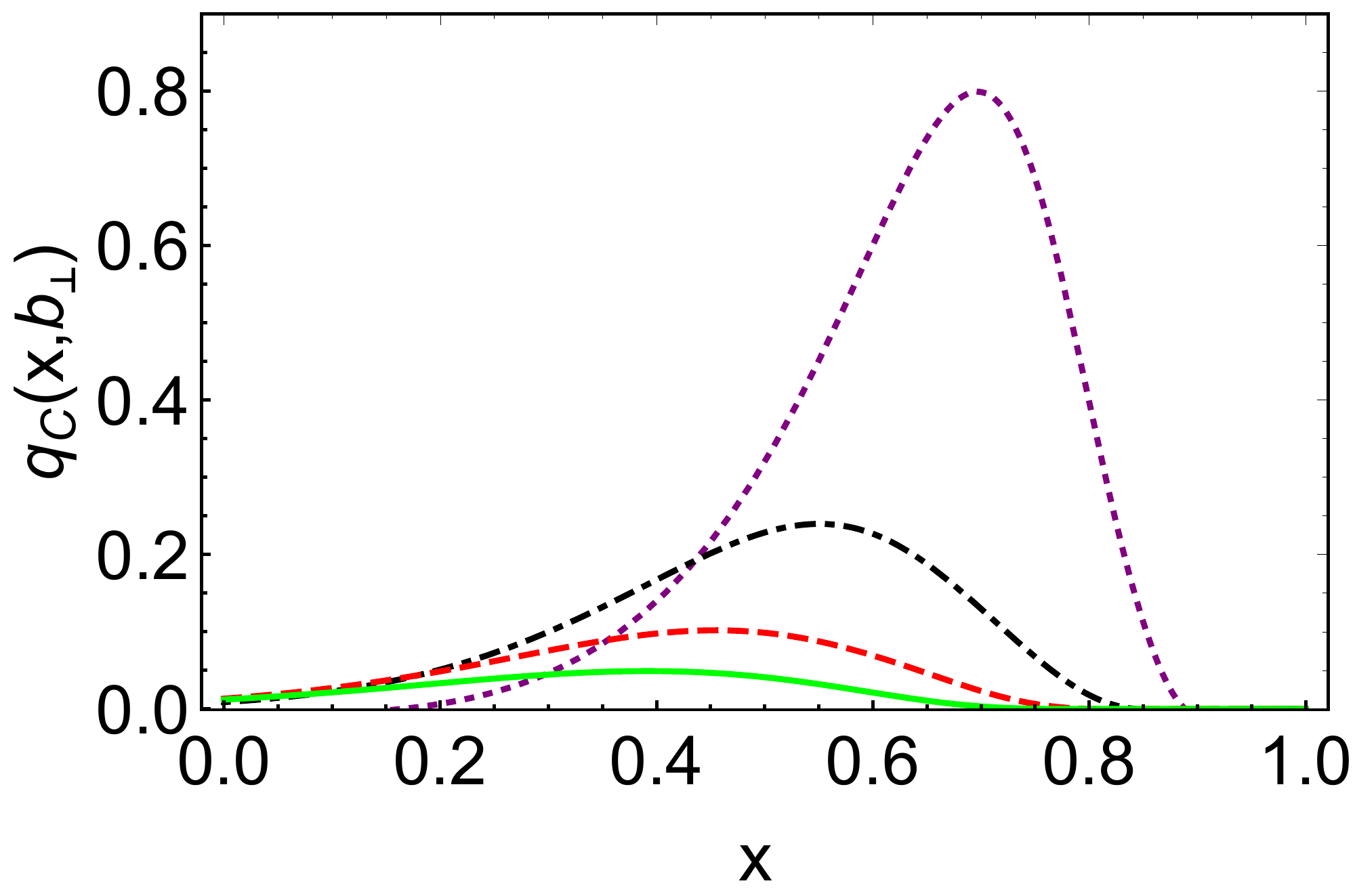}
\caption{The impact parameter dependent PDF $q_C\left(x,\bm{b}_{\perp}\right)$ with different $|\bm{b}_{\perp}|$; $0.15$ fm (dotted purple),
$0.25$ fm (dot-dashed black), $0.35$ fm (dashed red), $0.45$ fm (solid green).}\label{qc}
\end{figure}

The impact parameter dependent PDFs are defined as
\begin{align}\label{aG9}
q\left(x,b_{\perp}\right)&=\int \frac{\mathrm{d}^2\bm{\Delta}_{\perp}}{(2 \pi )^2}e^{-i\bm{b}_{\perp}\cdot \bm{\Delta}_{\perp}}H\left(x,0,-\bm{\Delta}_{\perp}^2\right),
\end{align}
which means the impact parameter dependent PDFs are the Fourier transform of GPDs at $\xi=0$. We obtain
\begin{align}\label{aG9}
&H_1\left(x,0,-\bm{\Delta}_{\perp}^2\right)=\frac{N_cZ_{\rho}}{4\pi^2} \bar{\mathcal{C}}_1(\sigma_6)-\frac{N_c Z_{\rho }}{2\pi^2}\int _0^{1-x} \mathrm{d}\alpha \,x   \bar{\mathcal{C}}_1(\sigma_8)\nonumber\\
&+\frac{N_c Z_{\rho }}{4\pi^2}\int _0^{1-x} \mathrm{d}\alpha\,(2xm_{\rho}^2+\bm{\Delta}_{\perp}^2(x-1))  \frac{\bar{\mathcal{C}}_2(\sigma_8)}{\sigma_8},
\end{align}
\begin{align}\label{aG9}
&H_2\left(x,0,-\bm{\Delta}_{\perp}^2\right)=\frac{N_cZ_{\rho}}{4\pi^2} \bar{\mathcal{C}}_1(\sigma_6)\nonumber\\
&- \frac{ N_c Z_{\rho }}{2 \pi^2}\int _0^{1-x} \mathrm{d}\alpha\, (1-x) \,\bar{\mathcal{C}}_1(\sigma_8)\nonumber\\
&+ \frac{ N_c Z_{\rho }}{4 \pi^2}\int _0^{1-x} \mathrm{d}\alpha\, (2m_{\rho}^2-(1-x)\bm{\Delta}_{\perp}^2) \frac{\bar{\mathcal{C}}_2(\sigma_8)}{\sigma_8},
\end{align}
\begin{align}\label{aG9}
&H_3\left(x,0,-\bm{\Delta}_{\perp}^2\right)=m_{\rho}^2 \frac{N_c Z_{\rho }}{\pi^2}\int _0^{1-x} \mathrm{d}\alpha\nonumber\\
&\times (\left(1-2\alpha -2x\right)\left(1-2\alpha\right)x  +2x^2 -x) \frac{\bar{\mathcal{C}}_2(\sigma_8)}{\sigma_8},
\end{align}
then the related impact parameter dependent PDFs can be obtained,
\begin{widetext}
\begin{subequations}\label{ims}
\begin{align}
q_C\left(x,\bm{b}_{\perp}\right)&=\int \frac{\mathrm{d}^2\bm{\Delta}_{\perp}}{(2 \pi )^2}e^{-i\bm{b}_{\perp}\cdot \bm{\Delta}_{\perp}}H_1\left(x,0,-\bm{\Delta}_{\perp}^2\right)\nonumber\\
&+\int \frac{\mathrm{d}^2\bm{q}_{\perp}}{(2 \pi )^2}e^{-i\bm{b}_{\perp}\cdot \bm{\Delta}_{\perp}}\frac{2}{3} \frac{\bm{\Delta}_{\perp}^2}{4m_{\rho}^2} \left( H_1\left(x,0,-\bm{\Delta}_{\perp}^2\right)-H_2\left(x,0,-\bm{\Delta}_{\perp}^2\right)+(1+\frac{\bm{\Delta}_{\perp}^2}{4m_{\rho}^2})H_3\left(x,0,-\bm{\Delta}_{\perp}^2\right) \right)\,, \\
q_M\left(x,\bm{b}_{\perp}\right)&=\frac{1}{G_M(0)}\int \frac{\mathrm{d}^2\bm{\Delta}_{\perp}}{(2 \pi )^2}e^{-i\bm{b}_{\perp}\cdot \bm{\Delta}_{\perp}}H_2\left(x,0,-\bm{\Delta}_{\perp}^2\right)\,, \\
q_Q\left(x,\bm{b}_{\perp}\right)&=\frac{1}{G_Q(0)}\int \frac{\mathrm{d}^2\bm{\Delta}_{\perp}}{(2 \pi )^2}e^{-i\bm{b}_{\perp}\cdot \bm{\Delta}_{\perp}}\left( H_1\left(x,0,-\bm{\Delta}_{\perp}^2\right)-H_2\left(x,0,-\bm{\Delta}_{\perp}^2\right)+(1+\frac{\bm{\Delta}_{\perp}^2}{4m_{\rho}^2})H_3\left(x,0,-\bm{\Delta}_{\perp}^2\right) \right),
\end{align}
\end{subequations}
\end{widetext}
which are similar to the expressions in Ref.~\cite{Mondal:2017lph}. We illustrate the impact parameter dependent PDFs in Figs. \ref{qc}-\ref{qq},
from the figures we can find that, with the increasing of $|\bm{b}_{\perp}|$, the values of $q_i\left(x,\bm{b}_{\perp}\right)$ are decreasing.
For the same $|\bm{b}_{\perp}|$, $q_C\left(x,\bm{b}_{\perp}\right)$ has the maximum value. $q_M \left(x,\bm{b}_{\perp}\right)$ and
$q_Q\left(x,\bm{b}_{\perp}\right)$ do not have much difference. For the position of the peaks, $q_C\left(x,\bm{b}_{\perp}\right)$ is at the largest $x$, while $q_Q\left(x,\bm{b}_{\perp}\right)$ is at the smallest $x$.

%
\begin{figure}
\centering
\includegraphics[width=0.47\textwidth]{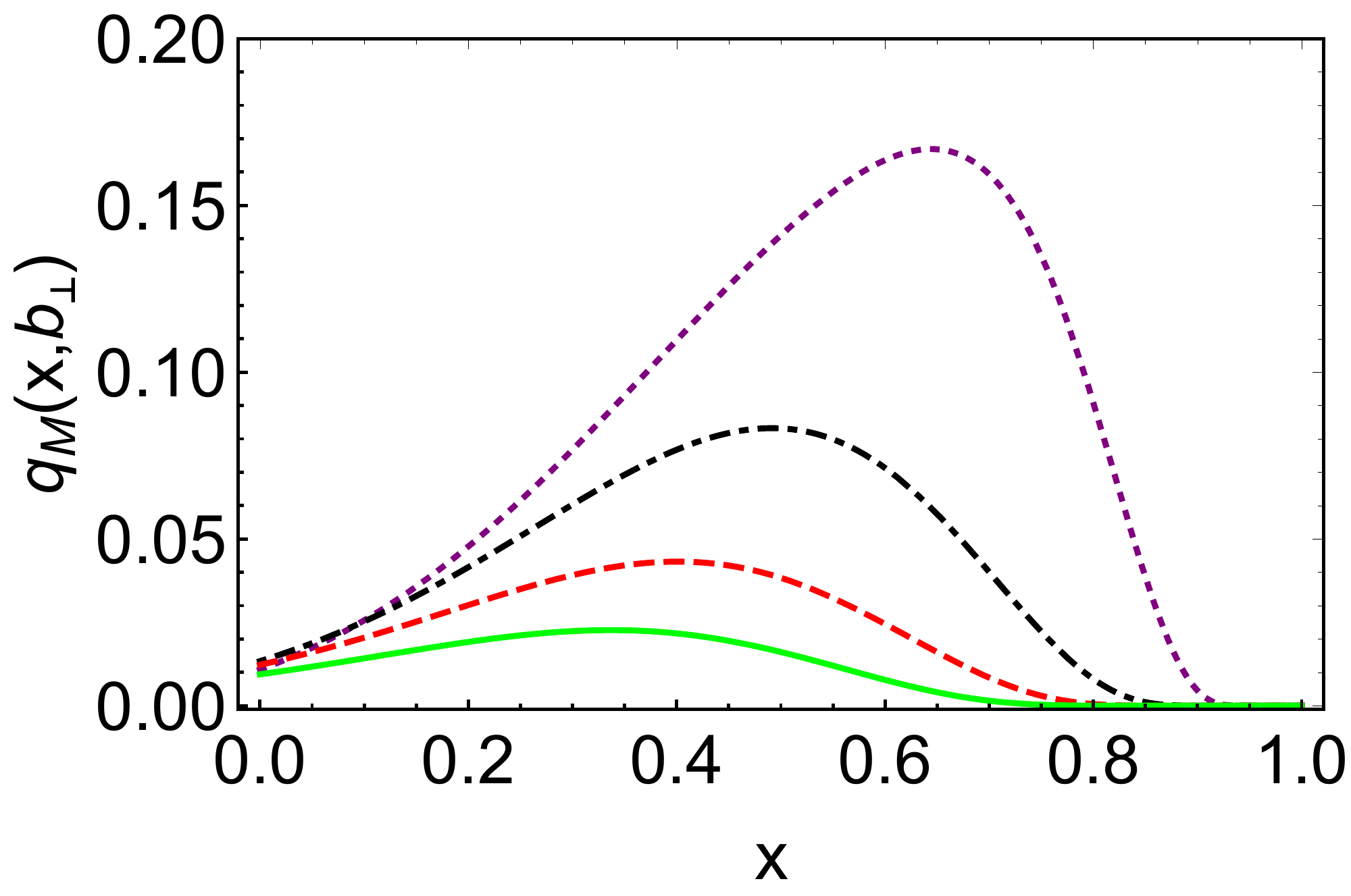}
\caption{The impact parameter dependent PDF $q_M\left(x,\bm{b}_{\perp}\right)$ with different $|\bm{b}_{\perp}|$; $0.15$ fm (dotted purple),
$0.25$ fm (dot-dashed black), $0.35$ fm (dashed red), $0.45$ fm (solid green).}\label{qm}
\end{figure}
\begin{figure}
\centering
\includegraphics[width=0.47\textwidth]{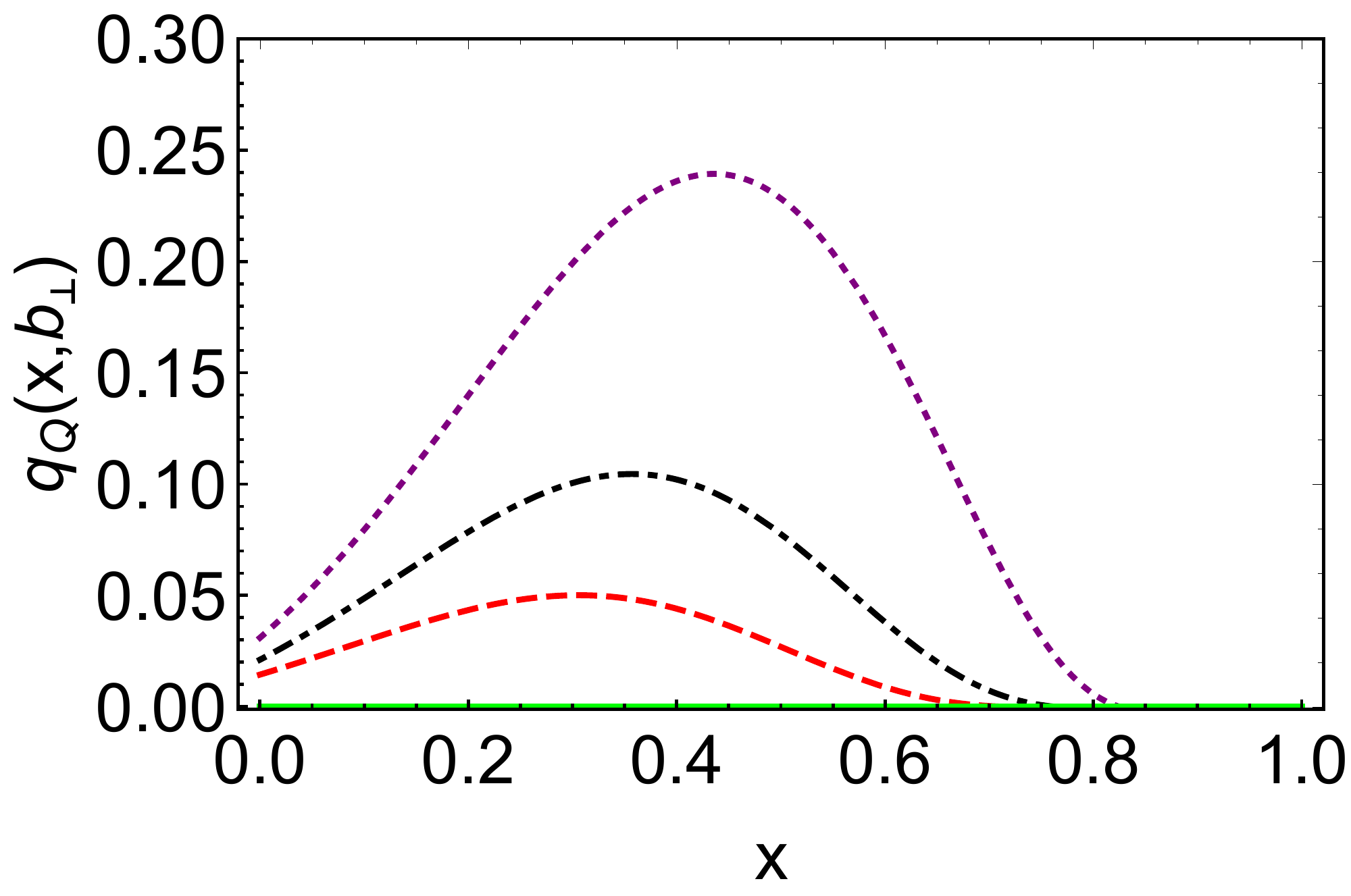}
\caption{The impact parameter dependent PDF $q_Q\left(x,\bm{b}_{\perp}\right)$ with different $|\bm{b}_{\perp}|$; $0.15$ fm (dotted purple),
$0.25$ fm (dot-dashed black), $0.35$ fm (dashed red), $0.45$ fm (solid green).}\label{qq}
\end{figure}

\section{Summary and conclusion}\label{excellent}
In this work, we evaluate the unpolarized and polarized $\rho$ meson generalized parton distributions (GPDs) in the framework of the
Nambu--Jona-Lasinio (NJL) model using proper time regularization. Based on the results, the properties of GPDs are checked, the results show that the required properties of GPDs are satisfied very well. Besides, the three independent distribution functions in deep inelastic scattering, $F_1^u(x)$, $b_1^u(x)$ and $g_1^u(x)$, and the Sachs-like charge, magnetic, and quadruple form factors $G_C(t)$, $G_M(t)$ and $G_Q(t)$, are presented. All the results in the present work
are generally consistent with those of previous calculations.

In the future, on one hand, we plan to discuss the gravitational form factors of $\rho$ meson, which contain the information on the spatial distributions of energy, spin, shear forces and pressure inside the system~\cite{Polyakov:2002yz}. The gravitational form factors are define through the matrix elements of the symmetric energy-momentum tensor, which can be obtained from the second Mellin moments of GPDs. The sum rules between
GPDs and energy-momentum tensor form factors can be found in Refs.~\cite{Abidin:2008ku,Taneja:2011sy}, which is an extension of our current research.
On the other hand, we will study the transversity GPDs, or helicity flip GPDs of $\rho$ meson introduced in Ref.~\cite{Diehl:2003ny}. As we mentioned, spin-$1$ particles have $9$ helicity conserving GPDs and $9$ helicity flip or transversity ones. Transversity GPDs appear as scalar functions in the decomposition of off-forward quark-quark correlators with a parton helicity flip. The helicity non-flip GPDs are chiral even, while the helicity flip GPDs are chiral-odd.

\acknowledgments
We would like to thank for constructive comments and technical assistance from Cédric Mezrag and Zhu-Fang Cui.
Work supported by: National Natural Science Foundation of China (under grant nos. 11775118).

\appendix
\section{Appendix 1: useful formulas}\label{AppendixT1}
Here we use the gamma-functions ($n\in \mathbb{Z}$, $n\geq 0$)
\begin{subequations}\label{cfun}
\begin{align}
\mathcal{C}_0(z)&:=\int_0^{\infty} \mathrm{d}s\, s \int_{\tau_{uv}^2}^{\tau_{ir}^2} \mathrm{d}\tau \, e^{-\tau (s+z)}\nonumber\\
&=z[\Gamma (-1,z\tau_{uv}^2 )-\Gamma (-1,z\tau_{ir}^2 )]\,, \\
\mathcal{C}_n(z)&:=(-)^n\frac{z^n}{n!}\frac{\mathrm{d}^n}{\mathrm{d}z^n}\mathcal{C}_0(z)\,, \\
\bar{\mathcal{C}}_i(z)&:=\frac{1}{z}\mathcal{C}_i(z),
\end{align}
\end{subequations}
where $\tau_{uv,ir}=1/\Lambda_{\text{UV},\text{IR}}$ are the infrared and ultraviolet regulators, respectively,
and $\Gamma (\alpha,y )$ is the incomplete gamma-function, $z$ represents the $\sigma$ functions in the following.

The $\sigma$ functions are define as
\begin{subequations}\label{cfun1}
\begin{align}
\sigma_1&=M^2-x(1-x)\Delta^2\,, \\
\sigma_2&=M^2-\frac{x+\xi}{1+\xi} \frac{1-x}{1+\xi} m_{\rho}^2\,, \\
\sigma_3&=M^2-\frac{x-\xi}{1-\xi}\frac{1-x}{1-\xi} m_{\rho}^2\,, \\
\sigma_4&=M^2-\frac{1}{4}(1+\frac{x}{ \xi })(1-\frac{x}{\xi }) t\,, \\
\sigma_5&=M^2-\alpha \left(1-\alpha \right)  m_{\rho}^2\nonumber\\
&-\left(\frac{\xi+x}{2\xi}-\alpha \frac{1+\xi}{2\xi}\right) (\frac{\xi-x}{2\xi}+\alpha \frac{1-\xi}{2\xi}) t\,, \\
\sigma_6&=M^2-x(1-x)m_{\rho}^2\,, \\
\sigma_7&=(x+y)(x+y-1)m_{\rho }^2-x\,y\,t+M^2\,, \\
\sigma_8&=M^2+(1-\alpha-x) \alpha \bm{\Delta}_{\perp}^2-x \left(1-x\right) m_{\rho}^2.
\end{align}
\end{subequations}

\bibliographystyle{apsrev4-1}
\bibliography{zhangrho}


\end{document}